\documentclass[aps,groupedaddress, superscriptaddress,showpacs,amsmath,amssymb,twocolumn,pra]{revtex4-1}
\usepackage{bm}
\usepackage[dvips]{graphicx}
\usepackage{wrapfig,subfigure}
\usepackage{amssymb}
\usepackage{color}
\usepackage[normalem]{ulem} \usepackage{soul,xcolor} \setstcolor{black} \setul{}{2.4pt}
\usepackage{amsmath,bm}
\usepackage{epstopdf}
\usepackage{mathtools} 
\usepackage{extarrows} 
 \usepackage{bbm}
\usepackage{stmaryrd}
\usepackage{graphicx}
\usepackage{hhline}
\usepackage{ulem}
\usepackage{chngcntr}

\def\be{\begin{equation}}
\def\ee{\end{equation}}
\newcommand{\nth}{\bar n_{\rm th}}

\newcommand{\ncoh}{\bar n_{\rm coh}}

\newcommand{\noise}{\xi_{\rm th}(t)}

\newcommand{\R}{{\rm Re}}

\begin{document}
\title{Measurement of small photon numbers in circuit QED resonators}
\author{Juan Atalaya}
\affiliation{Google Inc., 340 Main Street, Venice, CA 90291, USA}
\author{Alex Opremcak}
\affiliation{Google Inc., 340 Main Street, Venice, CA 90291, USA}
\author{Ani Nersisyan}
\affiliation{Google Inc., 340 Main Street, Venice, CA 90291, USA}
\author{Kenny Lee}
\affiliation{Google Inc., 340 Main Street, Venice, CA 90291, USA}
\author{Alexander N.\ Korotkov}
\affiliation{Google Inc., 340 Main Street, Venice, CA 90291, USA}
\affiliation{Department of Electrical and Computer Engineering, University of California, Riverside, CA 92521, USA}

\date{\today}

\begin{abstract}
Off-resonant interaction of fluctuating photons in a resonator with a qubit increases the qubit dephasing rate. We use this effect to measure a small average number of intracavity photons that are coherently or thermally driven. For spectral resolution, we do this by subjecting the qubit to a Carr-Purcell-Meiboom-Gill (CPMG) sequence and record the qubit dephasing rate for various periods between qubit $\pi$-pulses. The recorded data is then analyzed with formulas for the photon-induced dephasing rate that we have derived for the non-Gaussian noise regime with an arbitrary ratio $2\chi/\kappa$, where $2\chi$ is the qubit frequency shift due to a single photon and $\kappa$ is the resonator decay rate. We show that the presented CPMG dephasing rate formulas agree well with experimental results and demonstrate measurement of thermal and coherent photon populations at the level of a few $10^{-4}$.
\end{abstract}
\maketitle

{\it Introduction.---}In state-of-the-art circuit QED setups, the dispersive interaction~\cite{Blais2021} between qubits and residual photons in readout microwave cavities has been recognized as an important source of decoherence~\cite{Mooij2005, Schoelkopf2012, Rigetti2012, Oliver2016, Wellstood2017, Oliver2018, Devoret2019} that can prevent the $T_2$ time from reaching the no-pure-dephasing limit of $2T_1$ in flux~\cite{Oliver2016} and transmon~\cite{Devoret2019} qubits, even at the flux-insensitive point. Such residual photons can be {\it thermal} (due to, e.g., improperly filtered/attenuated blackbody radiation that impinges on the ports of the microwave cavity~\cite{Oliver2018, Devoret2019}) or {\it coherent} (due to, e.g., unintended driving of readout cavities), or both. To suppress but not eliminate photon-induced dephasing~\cite{Mooij2005}, dynamical decoupling (DD) techniques, such as the Carr-Purcell-Meiboom-Gill (CPMG) sequence~\cite{CPMG}, can be used~\cite{GoogleQAI2021}. Additionally, DD sequences have been used to probe the dephasing noise spectra~\cite{Cappellaro2017}. 

Characterization of very small intracavity photon numbers with superconducting qubits is of practical interest to the circuit QED community. Recent works have demonstrated measurement of a few $10^{-4}$ to a few $10^{-3}$  thermal photons on average~\cite{Devoret2019, Oliver2018, Oliver2016}. This measurement capability has been used in microwave radiometry~\cite{Devoret2021} and in recent improvements of microwave attenuators, aimed at thermalizing the microwave cavities at the mixing chamber temperature~\cite{Wellstood2017,Devoret2019}. Two experimental methods have been previously used to characterize the average thermal or coherent intracavity photon number ($\bar n_{\rm th}$ or $\bar n_{\rm coh}$). The first method uses a superconducting qubit subject to a Ramsey or spin-echo  sequence~\cite{Devoret2021, Devoret2019}. An upper bound for $\bar n_{\rm th}$ can be inferred by assuming that the measured pure dephasing rate is only due to thermal photon shot noise~\cite{Dykman1987, Clerk2007}. It has been reported that this method can detect thermal photon numbers of the order of a few $ 10^{-4}$~\cite{Devoret2019, Oliver2018}; however, it cannot distinguish between thermal and coherent photon populations, and it can be applied to tunable qubits only at the flux-insensitive point. The second method uses a spin locking protocol that probes the photon shot-noise spectrum~\cite{Oliver2018, Oliver2016}. The main feature of this method is that it can distinguish between thermal and coherent photon populations since the spectral line of the photon shot noise has a half-width of $\kappa$ for thermal photons and $\kappa/2$ for coherent photons~\cite{Gambetta2006, Oliver2018}, where $\kappa$ is the cavity mode decay rate. The spin locking method has been used to measure average thermal photon numbers of $6\times 10^{-3}$ {and below} \cite{Oliver2018}. 

\begin{figure}[!t]
    \centering
    \includegraphics[width=0.95\columnwidth, trim=0cm 1.cm 0cm 0cm, clip=True]{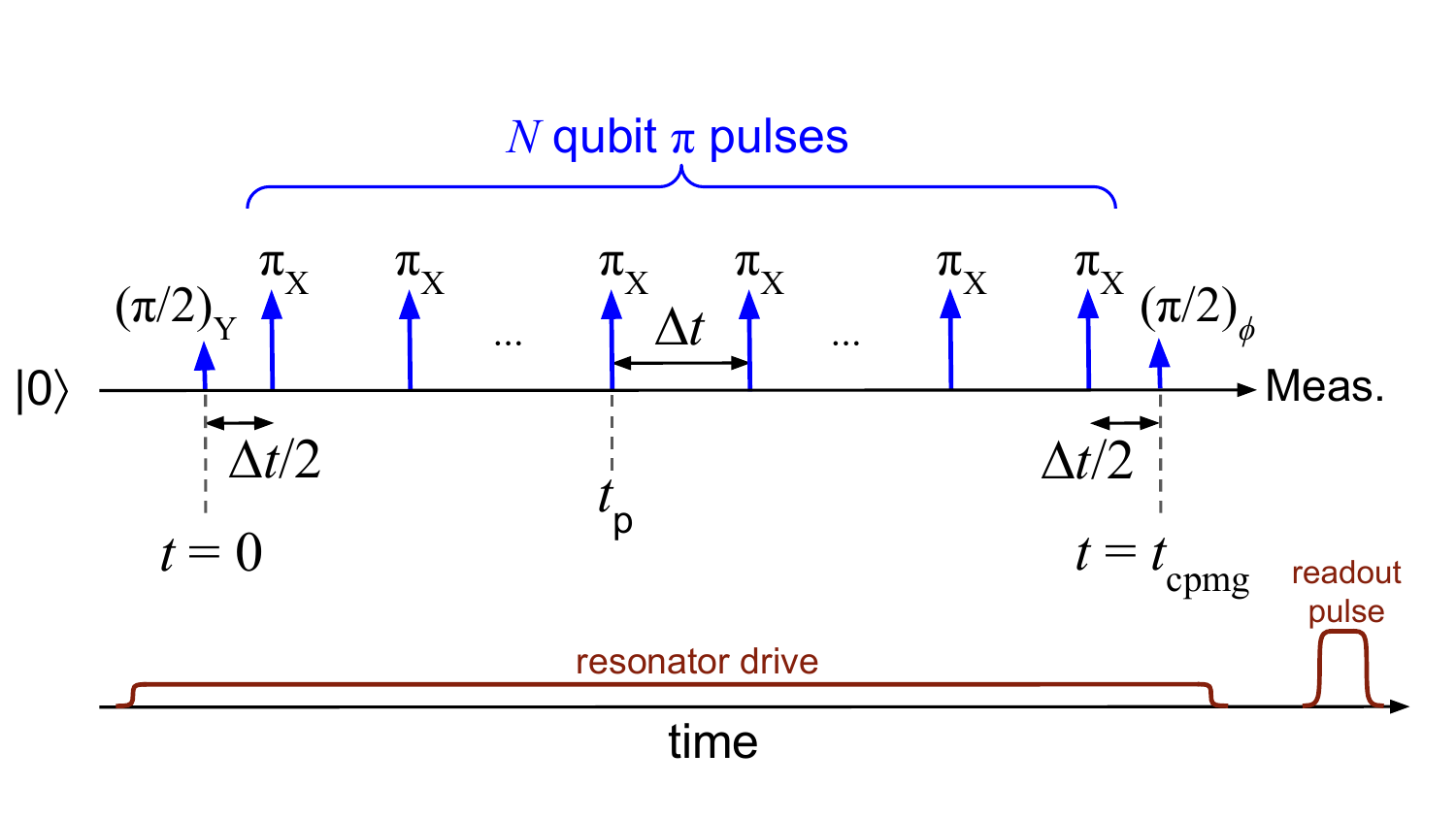}
    \caption{Experimental procedure. The qubit is subject to a CPMG sequence with $N$ qubit $\pi$-pulses, separated by the interpulse period $\Delta t$. At the same time, the resonator is continuously driven with a microwave tone of constant amplitude or by effectively white noise in order to add coherent or thermal photons, respectively (the resonator reaches steady state before the CPMG sequence begins at $t=0$). The sequence duration is $t_{\rm cpmg}=N\Delta t$  and $\phi$ indicates the microwave phase of the second $\pi/2$-pulse.} \label{CPMG_sequence}
\end{figure}

In this work, we present a method for measuring an ultrasmall average photon number in a resonator that is based on the dephasing rate of a qubit subject to a CPMG sequence. We demonstrate this metrology technique and measure {intentionally} added average thermal and coherent photon numbers at the level of $5\times 10^{-4}$. We also show that the ambient thermal photon number of our resonators is below $2\times 10^{-4}$. In this method, the qubit's CPMG dephasing rate, $\Gamma_2^{\rm cpmg}$, is measured for various periods $\Delta t$ between consecutive qubit $\pi$-pulses of the CPMG sequence (see Fig.~\ref{CPMG_sequence}). The specific dependence of $\Gamma_2^{\rm cpmg}$ on $\Delta t$ allows us to find contributions to the qubit dephasing from thermal or coherent intracavity photons and to infer the corresponding average photon numbers $\bar n_{\rm th}$ and $\bar n_{\rm coh}$. For this purpose we have derived analytical formulas for the photon-induced dephasing rate in CPMG, which are applicable for an arbitrary ratio between the dispersive shift $2\chi$ and resonator decay rate $\kappa$. Since the noise is significantly non-Gaussian when $2\chi/\kappa$ is not small, the standard filter-function approach is not applicable (as shown later). 
The derived formulas are also useful to calculate the reduction factor of the photon-induced dephasing rate by periodic DD sequences such as the CPMG and XY-4 in, e.g., quantum error correction applications~\cite{GoogleQAI2023}. The formulas show good agreement with numerical results and experimental data. 

 {\it Photon-induced dephasing in CPMG.}---We consider a qubit subject to a conventional CPMG sequence that includes $N$  $\pi$-pulses, separated by the period $\Delta t$. It also includes two $\pi/2$-pulses at the beginning and at the end of the sequence (see  Fig.~\ref{CPMG_sequence}) and has a duration $t_{\rm cpmg}=N\Delta t$. During the execution of the CPMG sequence, fluctuating photons in a microwave cavity (the qubit readout resonator in our experiments) induce additional dephasing onto the qubit due to dispersive interaction Hamiltonian $H_{\rm int}=-\chi \sigma_z \hat{n}$, where $\sigma_z=|0\rangle\langle0| - |1\rangle\langle1|$ acts on the qubit and $\hat n $ is the resonator number operator.

We focus on the qubit coherence, defined as  
\begin{align}
\label{coh-def}
\mathcal{C}(t_{\rm cpmg})=2|{\rm Tr}_{\rm res}\,{\rho}_{01}(t_{\rm cpmg})|,
\end{align}
where ${\rho}_{01}=\langle 0|  \rho |1\rangle$ is an operator in the resonator Hilbert space, $ \rho$ is the density matrix of the qubit-resonator system, and the trace is over the resonator degrees of freedom. The qubit coherence is evaluated at the moment $t_{\rm cpmg}$ immediately before the second $\pi/2$-pulse. Experimentally, the coherence is obtained from the qubit population difference measured after the second $\pi/2$-pulse with six equidistant values of the microwave phase $\phi$, which is then fitted sinusoidally as a function of $\phi$, so that $\mathcal{C}$ is the visibility (amplitude) of the oscillation.
In the absence of decoherence, $\mathcal{C}=1$. For analytics, let us assume that the only source of qubit decoherence is dephasing from photon shot noise and that the qubit pulses are ideal and instantaneous.

Using the standard filter-function approach~\cite{Cappellaro2017,DasSarma2008,Hirayama2011,Suter2011} based on the Gaussian approximation, for  $N \gg 1$ and dephasing due to thermal photons in the resonator,  
we obtain the exponential decay of $\mathcal{C}$ with  dephasing rate 
\begin{align}
\label{Gamma_phi_filter_func}
    \Gamma_\varphi^{\rm th,\,ff} &= \frac{4\chi^2 \bar n_{\rm th}}{\kappa} \left(1 - \frac{\tanh(\kappa\Delta t/2)}{\kappa\Delta t/2} \right) ,  
\end{align}
while for {\it resonant} coherent photons, $\kappa$ in this formula should be replaced by $\kappa/2$ and the average photon number $\bar n_{\rm th}$ replaced by $\bar n_{\rm coh}$ -- see  Supplemental Material (SM)~\cite{SM}. These formulas assume $|2\chi | \ll \kappa$ and, as shown later, are very inaccurate in the experimentally relevant regime of moderate/strong dispersive coupling ($|2\chi | \agt \kappa$), where the qubit frequency noise cannot be approximated by a Gaussian process. (Gaussian noise usually implies many small fluctuations within the decoherence time, which is not the case here.) An attempt to introduce the phenomenological factor $[1+(2\chi/\kappa)^2]^{-1}$ to reproduce the low-frequency dephasing \cite{Dykman1987, Clerk2007,Gambetta2006}, helps only for $\kappa \Delta t \gg 1$.

Instead of using the filter-function approach, we describe the coherence evolution $\mathcal{C}(t)$ in the way that is rigorously derived in the SM \cite{SM} but can be understood physically \cite{Korotkov2016} as replacing the qubit state $(|0\rangle+|1\rangle)/\sqrt{2}$ after the first $\pi/2$-pulse with either $|0\rangle$ or $|1\rangle$. Then there  is no qubit-resonator entanglement, and the resonator field evolves as either $\alpha_0(t)$ or $\alpha_1(t)$, given by simple evolution equations. The coherence $\mathcal{C}$ then can be found via quantum overlap of the corresponding leaked resonator fields. Thus, the qubit coherence is~\cite{Gambetta2006, Korotkov2016,SM}
\begin{align}
\label{coh-formula}
    \mathcal{C}(t_{\rm cpmg}) = \left|\left\langle\exp\left[\int_0^{t_{\rm cpmg}}{\rm d}t\, 2i\tilde{\chi}(t) \alpha_0(t)\alpha^*_1(t)\right]\right\rangle\right|,
\end{align}
where $\tilde{\chi}(t)=\pm\chi$ is a piecewise constant function that flips sign after each $\pi$-pulse, and the complex-valued dynamical variables $\alpha_0(t)$ and $\alpha_1(t)$ evolve as
\begin{eqnarray}
   && \dot \alpha_{0,1}(t) = -\tilde\gamma_{0,1}(t) \alpha_{0,1} + \sqrt{\kappa}\, \noise + \sqrt{\kappa}\,F_{\rm d}(t) , \label{EOM-alpha0-1}
    \\
     &&   \tilde\gamma_{q}(t)=\kappa/2-i [\delta \omega_{\rm d} + (-1)^q \tilde\chi(t)], \,\,\, q=0,1 . 
        \label{gamma0-1}
\end{eqnarray}
Here $F_{\rm d}(t)$ is the complex amplitude of the coherent resonator  drive with frequency $\omega_{\rm d}$ (which defines the rotating frame), $\delta\omega_{\rm d}=\omega_{\rm d}- \omega_{\rm res}$ is the detuning, and  $\omega_{\rm res}$ is the average of the resonator frequencies when the qubit is in state $|0\rangle$ or $|1\rangle$. Note that the decay factors $\tilde\gamma_{0}(t)$ and $\tilde\gamma_{1}(t)$ are complex-valued piecewise constant functions, which include detunings for the two paths of resonator evolution. 
The intracavity thermal population is driven by a complex-valued Gaussian white noise $\noise$,
\begin{align}
\label{white_noise_corr}
    \langle \xi^*_{\rm th}(t) \, \xi_{\rm th}(t') \rangle = \bar n_{\rm th}\delta (t-t'), \,\,\,  \langle \noise\rangle = 0,
\end{align}
where $\bar{n}_{\rm th}$ is the average intracavity thermal photon number; $\nth =[\exp(\hbar \omega_{\rm res}/k_{\rm B}T_{\rm res})-1]^{-1}$ in the case of thermal equilibrium with resonator temperature $T_{\rm res}$. The notation $\langle \cdot\rangle$ indicates averaging over noise realizations.  We assume  $\alpha_0(0)=\alpha_1(0)=0$, though this is not important. If there is no coherent drive, then $F_{\rm d}=0$ and $\delta\omega_{\rm d}=0$. 
We point out that Eqs.~\eqref{coh-formula}--\eqref{white_noise_corr} are also valid for DD sequences with $\pi$-pulses applied at arbitrary times~\cite{Uhrig-2007}. 

We consider the experimentally relevant regime where the qubit coherence $\mathcal{C}$ decays exponentially with $N$ (and therefore sequence duration $t_{\rm cpmg}$), while still not being too small. As shown in the SM~\cite{SM}, this occurs for CPMG sequences longer than $\sim 3\kappa^{-1}$ (or $\sim 6\kappa^{-1}$) for dephasing due to thermal (or coherent) photons, with a sufficiently small average photon number. Then evolution of $\alpha_0(t)$ and $\alpha_1(t)$ reaches a quasi-steady-state regime, in which all averages are practically periodic with period $2\Delta t$ and the roles of $\alpha_0(t)$ and $\alpha_1(t)$ are exchanged after $\Delta t$. Only in this regime we can introduce the photon-induced qubit dephasing rate $\Gamma_\varphi$.

We first discuss photon-induced dephasing due to thermal photons, focusing on the limit of small average photon numbers, $\nth\ll1$. In this limit, using Eq.~\eqref{coh-formula}, the photon-induced dephasing rate can be approximated as 
\begin{align}
\label{Gamma-phi-approx}
       & \Gamma_\varphi^{\rm th}  =  -{\rm Re} \left[  \int_{t_{\rm p}}^{t_{\rm p}+\Delta t} {\rm d}t\, 2i\tilde\chi(t) \mathcal{A}(t)\right] /\Delta t, 
    \\ 
    \label{Ath}
     & \mathcal{A}(t) \equiv  \langle\alpha_0(t)\alpha_1^*(t)\rangle,
\end{align}
where $t_{\rm p}$ indicates the moment of one (any) of the $\pi$-pulses after reaching the quasi-steady-state regime. Since in this regime $\alpha_0$ and $\alpha_1$ exchange their roles after $\Delta t$, we can use approximation 
\begin{align}
\label{A-condition}
    \mathcal{A}(t+\Delta t) = \mathcal{A}^*(t).  
\end{align}
Note that  Eq.~\eqref{Gamma-phi-approx} gives only the leading-order term of $\Gamma_\varphi^{\rm th}$ dependence on $\nth$; a more general result is given in the SM~\cite{SM} that also applies to moderate values of $\nth$.

Without loss of generality, let us assume that $\tilde \chi(t) = \chi$ within the interval $t\in(t_{\rm p}, t_{\rm p}+\Delta t)$. Using Eqs.~\eqref{EOM-alpha0-1}--\eqref{white_noise_corr}, it is straightforward to show that $\mathcal{A}(t)$ evolves as 
\begin{align}
\label{EOM-At}
    \dot{\mathcal{A}}(t) = -(\kappa - 2\chi i)\mathcal{A} + \kappa\bar n_{\rm th}.
\end{align}
Solving this equation and using the condition $\mathcal{A}(t_{\rm p}+\Delta t) = \mathcal{A}^*(t_{\rm p})$ following from Eq.~\eqref{A-condition}, we find 
\begin{align}
\label{At_sol}
    & \mathcal{A}(t) = e^{-\kappa_-  (t-t_{\rm p})}\left[\mathcal{A}(t_{\rm p}) - \frac{\kappa\bar n_{\rm th}}{\kappa_-}\right] + \frac{\kappa\bar n_{\rm th}}{\kappa_-}, 
    \\
    & \label{A0}
    \mathcal{A}(t_{\rm p}) = \frac{\kappa \bar n_{\rm th}}{\kappa_-} - i\kappa\bar n_{\rm th}\frac{{\rm Im}[\kappa_- ^{-1}]\left(1 - e^{-\kappa_- \Delta t}\right)^*}{\sinh(\kappa\Delta t)e^{-\kappa\Delta t} },
\end{align}
where $\kappa_- \equiv \kappa - 2\chi i$.  Using this solution for $\mathcal{A}(t)$, we evaluate  the integral in Eq.~\eqref{Gamma-phi-approx} with $\tilde\chi(t)=\chi$ and obtain the thermal-photon-induced dephasing rate ($\nth\ll 1$),
\begin{align}
\label{thermal-main-result-p1}
    & \Gamma_\varphi^{\rm th}(\Delta t, \nth) = \frac{4\chi^2\, \nth}{\kappa\left[1+(2\chi/\kappa)^2\right]} \, {\mathcal{R}}_{\rm th}(\Delta t),
    \\
    \label{thermal-main-result-p2}
    & \mathcal{R}_{\rm th}(\Delta t) = 1 - \frac{\cosh(\kappa\Delta t) - \cos(2\chi\Delta t)}{(\kappa\Delta t/2)[1+(2\chi/\kappa)^2]\sinh(\kappa\Delta t)}, 
\end{align}
where the first term in Eq.~\eqref{thermal-main-result-p1} is the low-frequency limit~($\Delta t\to \infty$) of the qubit dephasing rate, which  agrees with Eqs.~(11)--(12) of Ref.~\cite{Dykman1987} and with Eq.~(44) of Ref.~\cite{Clerk2007}, while  $\mathcal{R}_{\rm th}$ is the reduction factor ($0\leq \mathcal{R}_{\rm th} \leq 1$) due to the CPMG sequence with interpulse period $\Delta t$ (note that  $\mathcal{R}_{\rm th}\to 1$ as $\Delta t \to \infty$ and $\mathcal{R}_{\rm th}\to 0$ as $\Delta t \to 0$). Equations~\eqref{thermal-main-result-p1}--\eqref{thermal-main-result-p2} is our main result for the dephasing rate from thermal photons; it holds for an arbitrary ratio $2\chi/\kappa$, reducing to Eq.~(\ref{Gamma_phi_filter_func}) when $2\chi/\kappa \ll 1$.

We now discuss dephasing due to coherent photons. To describe it, we  use Eqs.~\eqref{coh-formula}--\eqref{gamma0-1} without the noise term, $\xi_{\rm th}=0$, and without averaging in Eq.~\eqref{coh-formula}. We assume a constant coherent drive amplitude, $F_{\rm d}(t)=F_{\rm d}^{\rm dc}$. 

In the regime of exponential decay of the coherence, the photon-induced dephasing rate $\Gamma^{\rm coh}_\varphi$ from coherent photons can still be obtained from Eq.~\eqref{Gamma-phi-approx}, in which $\mathcal{A}(t)=\alpha_0 (t)\alpha_1^*(t)$ no longer requires averaging, but the time interval $(t_{\rm p},t_{\rm p}+\Delta t)$ should still assume reaching the quasi-steady-state regime, for which we can use approximation 
\begin{align}
\label{alpha-cond-coh-case}
    \alpha_0(t+\Delta t) = \alpha_1(t), \;\; \alpha_1(t+\Delta t) = \alpha_0(t). 
\end{align}
Note that while for thermal photons Eq.~\eqref{Gamma-phi-approx} is valid only in the limit $\nth\ll 1$, for the coherent drive we can consider  larger average photon numbers $\ncoh$ as long as the resulting dephasing rate is sufficiently small: $\Gamma^{\rm coh}_\varphi\ll {\rm min}(1/\Delta t, \kappa)$.

Proceeding as in the case of thermal photons, we solve Eq.~\eqref{EOM-alpha0-1} to obtain the trajectories $\alpha_0(t)$ and $\alpha_1(t)$ within the time interval $(t_{\rm p}, t_{\rm p}+\Delta t)$ and use the condition Eq.~\eqref{alpha-cond-coh-case} with $t=t_{\rm p}$ to find the initial values. Substituting the result into Eq.~\eqref{Gamma-phi-approx} and calculating the integral, we obtain the dephasing rate $\Gamma_\varphi^{\rm coh}$ due to fluctuating coherent photons in the resonator. In the case of no detuning, $\delta\omega_{\rm d}=0$, the final result is  
\begin{align}
\label{coh-main-result-p1}
    & \Gamma_\varphi^{\rm coh}(\Delta t, \ncoh)  = \frac{8\chi^2 \ncoh}{\kappa[1+(2\chi/\kappa)^2]} \, \mathcal{R}_{\rm coh}(\Delta t),
    \\ 
    \label{coh-main-result-p2}
    & \mathcal{R}_{\rm coh}(\Delta t) =\;1 - \frac{\cosh(\kappa\Delta t/2)-\cos(\chi\Delta t)}{(\kappa\Delta t/4)\sinh(\kappa\Delta t/2)[1+(2\chi/\kappa)^2]} \nonumber \\
    & \hspace{2.0cm} \times \left[1+\frac{\chi}{\kappa}\frac{\sin(\chi\Delta t)}{\sinh(\kappa\Delta t/2)}-2\left(\frac{\chi}{\kappa}\right)^2\right], 
\end{align}
where $\ncoh = \kappa|F^{\rm dc}_{\rm d}|^2/[(\kappa/2)^2+\chi^2]$ is the average intracavity photon number for both qubit states $|0\rangle$ and $|1\rangle$ (without CPMG), while the factor $\mathcal{R}_{\rm coh}(\Delta t)$ is due to the CPMG sequence with  period $\Delta t$. In the limit $\Delta t\to \infty$, we get $\mathcal{R}_{\rm coh} \to 1$ and our result for $\Gamma_\varphi^{\rm coh}$ agrees with the dephasing rate  given by Eq.~(69) of Ref.~\cite{Clerk2007}, Eq.~(5.20) of Ref.~\cite{Gambetta2006}, and Eq.~(17) of Ref.~\cite{Korotkov2015}.  When $\Delta t\to 0$, we get $\mathcal{R}_{\rm coh} \to 0$. Note that $\mathcal{R}_{\rm coh}$ can exceed 1 when $2\chi/\kappa > 1.393$ and $(2\chi/2\pi)\Delta t$ is near an odd integer. 
In the case of a non-zero detuning $\delta \omega_{\rm d}$, the formula for $\Gamma_\varphi^{\rm coh}(\Delta t, \ncoh)$ is lengthy; it is given in the  SM~\cite{SM}.

While the analytical results \eqref{thermal-main-result-p1}--\eqref{thermal-main-result-p2} and \eqref{coh-main-result-p1}--\eqref{coh-main-result-p2} assume instantaneous $\pi$-pulses, they agree well (somewhat surprisingly) with numerical simulations even when pulse duration occupies up to half of $\Delta t$ and is comparable to $1/\kappa$ \cite{SM}. We emphasize that the results for $\Gamma_\varphi^{\rm coh}$ and $\Gamma_\varphi^{\rm th}$ hold for an arbitrary ratio  $2\chi/\kappa$, in contrast to the filter-function result, Eq.~\eqref{Gamma_phi_filter_func}, which is valid only if  $2\chi/\kappa \ll 1$.

{\it Experimental results.---}To benchmark our formulas and to measure photon-induced dephasing rate due to thermal and coherent photons in a resonator with metrological precision, we have performed experiments using the procedure depicted in Fig.~\ref{CPMG_sequence}. Our experiments are conducted in the moderate dispersive coupling regime with a ratio $2\chi/\kappa \approx 0.7$, where the resonator decay rate is $\kappa=(19.4\,{\rm ns})^{-1}=2\pi\times 8.2$~MHz while the dispersive coupling parameter is $2\chi = 2\pi\times 5.7$~MHz. The qubit pulses have a duration of 25~ns which limits the minimum period $\Delta t$ of $\pi$-pulses (we use $\Delta t \geq 40$~ns, so that the CPMG sequence frequency $f_s = 1/2\Delta t$ is limited by 12.5 MHz). We use a tunable transmon with frequency $\omega_{\rm q} = 2\pi\times 4.1$~GHz (nonlinearity is 229 MHz), set near the flux-insensitive point, and a resonator with frequency $\omega_{\rm res}=2\pi\times5.2$~GHz.

\begin{figure}[t!]
    \centering
    \includegraphics[width=0.95\columnwidth,trim=0cm 0cm 0cm 2cm,clip=True]{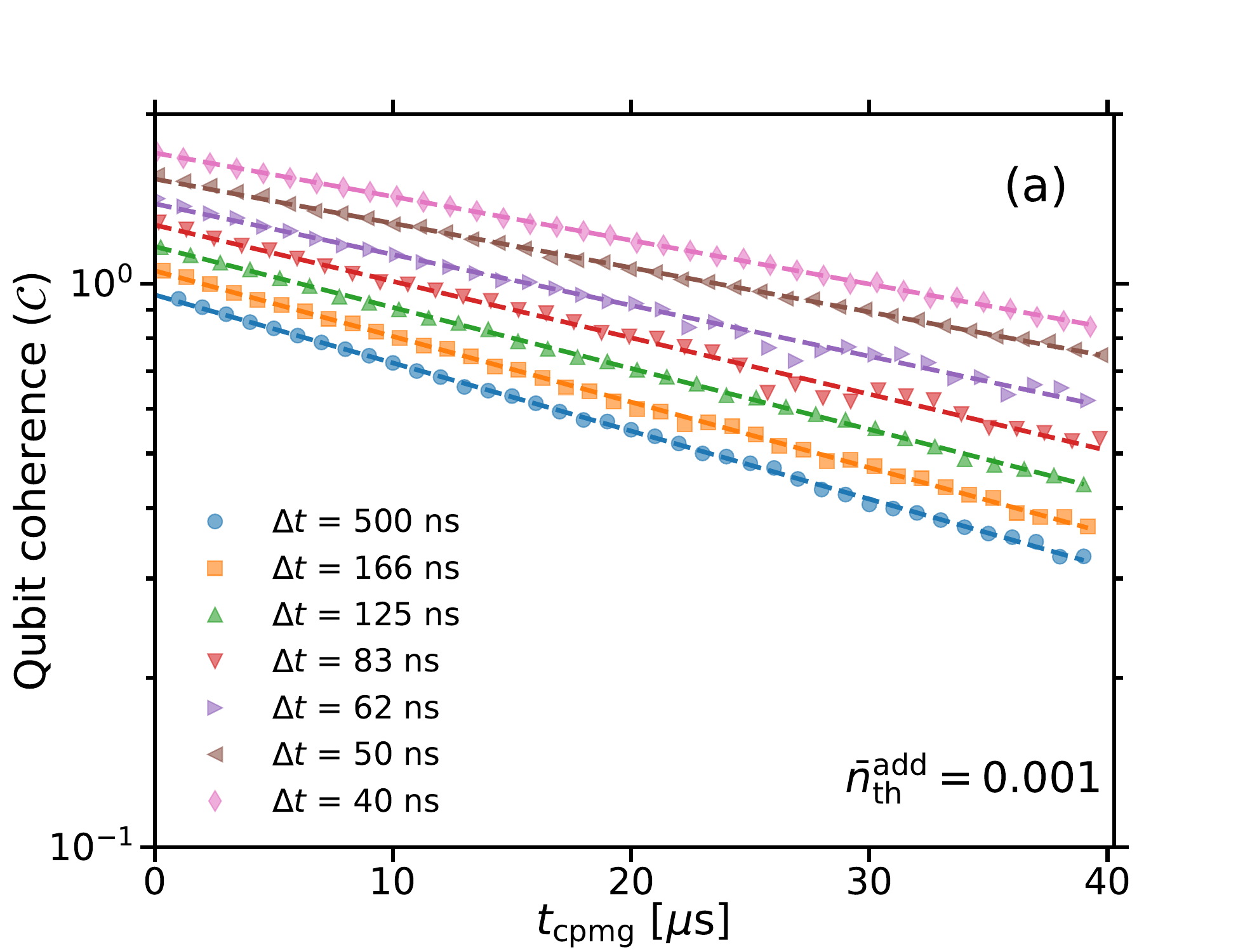}
    \includegraphics[width=0.95\columnwidth,trim=0cm 0cm 0cm 2cm,clip=True]{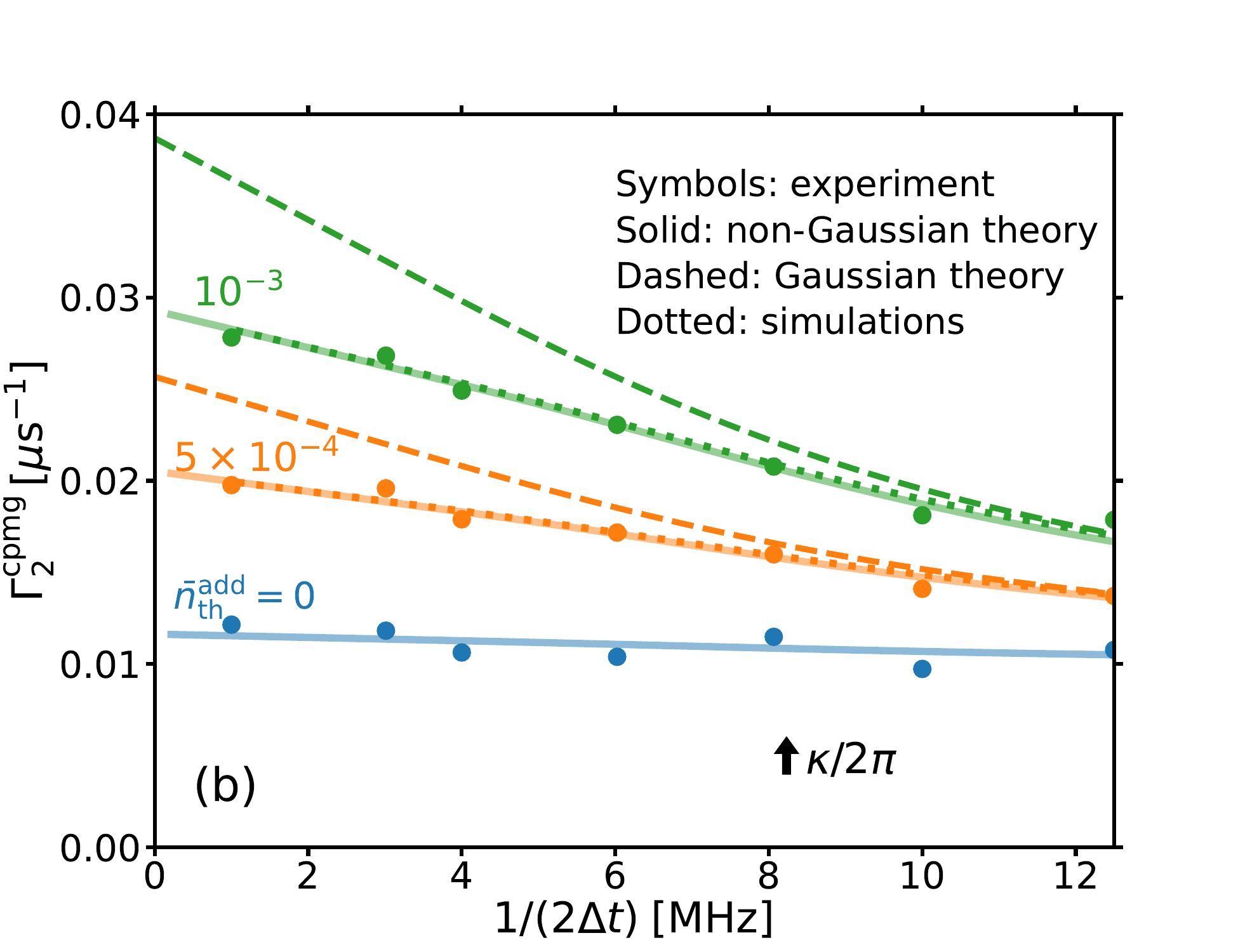}
    \caption{CPMG dephasing from thermal photon shot noise.
    Panel~(a) shows the coherence decay with CPMG time for several interpulse durations $\Delta t$ and for an added thermal population of  $\nth^{\rm add}=10^{-3}$. For visibility, we shift vertically the experimental points (symbols), except those for $\Delta t=500$~ns. The dephasing rates $\Gamma_2^{\rm cpmg}$ are extracted from exponential fits (dashed lines). Panel~(b) shows the CPMG dephasing rates as functions of the CPMG sequence frequency $f_s=1/2\Delta t$ for added thermal populations of $\nth^{\rm add}=0$ (blue), $5\times10^{-4}$ (orange) and $10^{-3}$ (green). Symbols indicate the experimental data and solid lines are fits based on Eqs.~\eqref{thermal-main-result-p1}--\eqref{thermal-main-result-p2} with a pedestal $\Delta \Gamma_2$. The dashed lines show the filter-function theory [Eq.~\eqref{Gamma_phi_filter_func}] and the dotted lines show numerical results that include the shape and duration (25~ns) of the experimental qubit pulses.}
    \label{fig:thermal_expt_results}
\end{figure}

Let us first discuss experimental results for qubit dephasing from thermal photons in the resonator. The ambient thermal population in the resonator is too small for a confident fit with our formula. Therefore, we controllably add thermal photons to the resonator by driving it with an engineered  broad-band noise as in Refs.~\cite{Devoret2019, Oliver2016, Oliver2018}. The correlation time of the applied noise is roughly 1~ns that is much shorter than the resonator decay time $\kappa^{-1}$, which sets the correlation time of the thermal photon number fluctuations inside the resonator. By varying the applied noise power $\mathcal{P}_{\rm drive}$, we vary the total average photon number, $\bar n_{\rm th} = \nth^{\rm add} + \bar n_{\rm th}^{\rm amb}$, where $\bar n_{\rm th}^{\rm amb}$ denotes the ambient thermal photon population in the resonator and $\nth^{\rm add} \propto \mathcal{P}_{\rm drive}$ denotes the added photon number. The proportionality coefficient between $\nth^{\rm add}$ and  $\mathcal{P}_{\rm drive}$ is calibrated by using spin-echo dephasing data, as discussed in the SM~\cite{SM}. 

Figure~\ref{fig:thermal_expt_results} shows experimental results for the CPMG dephasing due to thermal photons. By fixing $\Delta t$ and changing $N$, we find dephasing rate $\Gamma_2^{\rm cpmg}$ and then find its dependence on varying $\Delta t$.  Panel~(a) illustrates the exponential decay of the qubit coherence $\mathcal{C}$ as a function of the CPMG sequence time $t_{\rm cpmg}$ for an added thermal population of $\nth^{\rm add}=10^{-3}$ (for clarity, the data for different $\Delta t$ are shifted vertically). Similar results are also obtained for $\nth^{\rm add}=0$ and $5\times 10^{-4}$ (not shown). Each data point (symbol) in the Panel~(a) is obtained from averaging over 120,000 realizations, including 60 different realizations of the engineered noise. The number $N$ of qubit $\pi$-pulses is chosen to be $N\geq 2$ and we check that $\kappa N\Delta t \ge 4$ to avoid the initial non-exponential decay regime of the qubit coherence (ideally $N\gg 1$ and $\kappa N\Delta t \gg 1$); the chosen $N$ are even and correspond to $N\Delta t$ separation by about 1 $\mu$s. For the method it is important to have a good exponential fit of the dependence $\mathcal{C}(t_{\rm cpmg})$ -- see dashed lines in Fig.~\ref{fig:thermal_expt_results}(a). From these exponential fits, $\mathcal{C} \propto \exp(-\Gamma_2^{\rm cpmg} t_{\rm cpmg})$, we extract the experimental CPMG dephasing rates $\Gamma_2^{\rm cpmg}$. 

Figure~\ref{fig:thermal_expt_results}(b) shows the measured dephasing rate $\Gamma_2^{\rm cpmg}$ as a function of the CPMG sequence frequency $f_s=1/2\Delta t$ (in the filter-function approach, the noise at $f_s$ gives the main contribution to $\Gamma_2^{\rm cpmg}$ \cite{SM}) for added thermal photon populations $\nth^{\rm add} = 0$ (blue), $5\times 10^{-4}$ (orange) and $10^{-3}$ (green). The experimental data is indicated by symbols and the solid lines show the theoretical fit for the CPMG dephasing rate, 
\begin{align}    
\Gamma_2^{\rm cpmg} = \Gamma_\varphi^{\rm th}(\Delta t, \nth) + \Delta \Gamma_2,
\label{Gamma2_cpmg_fit}
\end{align} 
where $\Gamma_\varphi^{\rm th}$ is given by Eqs.~\eqref{thermal-main-result-p1}--\eqref{thermal-main-result-p2} and  $\Delta \Gamma_2$ is the dephasing rate contribution not related to photon shot noise, assumed to be independent of $\Delta t$. We fit all the experimental data points of Fig.~\ref{fig:thermal_expt_results}(b) using four fitting parameters: three values of $\nth$ and common $\Delta \Gamma_2$. Then we obtain $\Delta \Gamma_2 = (101.0\,\mu{\rm s})^{-1}$ and the photon populations  $\nth=(1.0\pm0.4)\times10^{-4}$, $(6.3\pm0.4)\times10^{-4}$, and $(1.15\pm0.04)\times10^{-3}$ for the blue, orange, and green lines, respectively. These values of $\nth$ are consistent with intended values of $\nth^{\rm add}$ and additional contribution from ambient population $\nth^{\rm amb}$ of crudely $1.3\times 10^{-4}$. 
If we fit only the blue points ($\nth^{\rm add}$=0), then we get $\nth^{\rm amb}=(1.5\pm0.8)\times10^{-4}$ and $\Delta \Gamma_2= (106.9 \, \mu {\rm s})^{-1}$. From these results we conclude that the ambient thermal photon number is below $2\times 10^{-4}$ and the inaccuracy of our method for $\nth^{\rm amb}$ is crudely $10^{-4}$. 

We emphasize a very good agreement in  Fig.~\ref{fig:thermal_expt_results}(b) between the experimental data and our analytical theory (orange and green solid lines). This is in sharp contrast with the Gaussian-approximation-based filter-function theory, Eq.~\eqref{Gamma_phi_filter_func}, depicted by the dashed lines, for which we use the same values of $\nth$ and $\Delta \Gamma_2$ that were obtained from the above fits. If in  Eq.~\eqref{Gamma_phi_filter_func} we replace $\nth$ with  $\nth/[1+(2\chi/\kappa)^2]$ to get agreement with Eq.~\eqref{thermal-main-result-p1} in the low-frequency case, $1/(2\Delta t)\to 0$, then there will be a significant disagreement for $1/(2\Delta t)$ comparable or bigger than $\kappa$. The dotted lines in Fig.~\ref{fig:thermal_expt_results}(b) show numerical results, which take into account the shape and duration of the $\pi$-pulses.
We see that these lines practically coincide with the solid lines, so our analytics works very well, only slightly underestimating the dephasing rate for short $\Delta t$ comparable with the pulse duration.

\begin{figure}
    \centering
    \includegraphics[width=0.95\columnwidth]{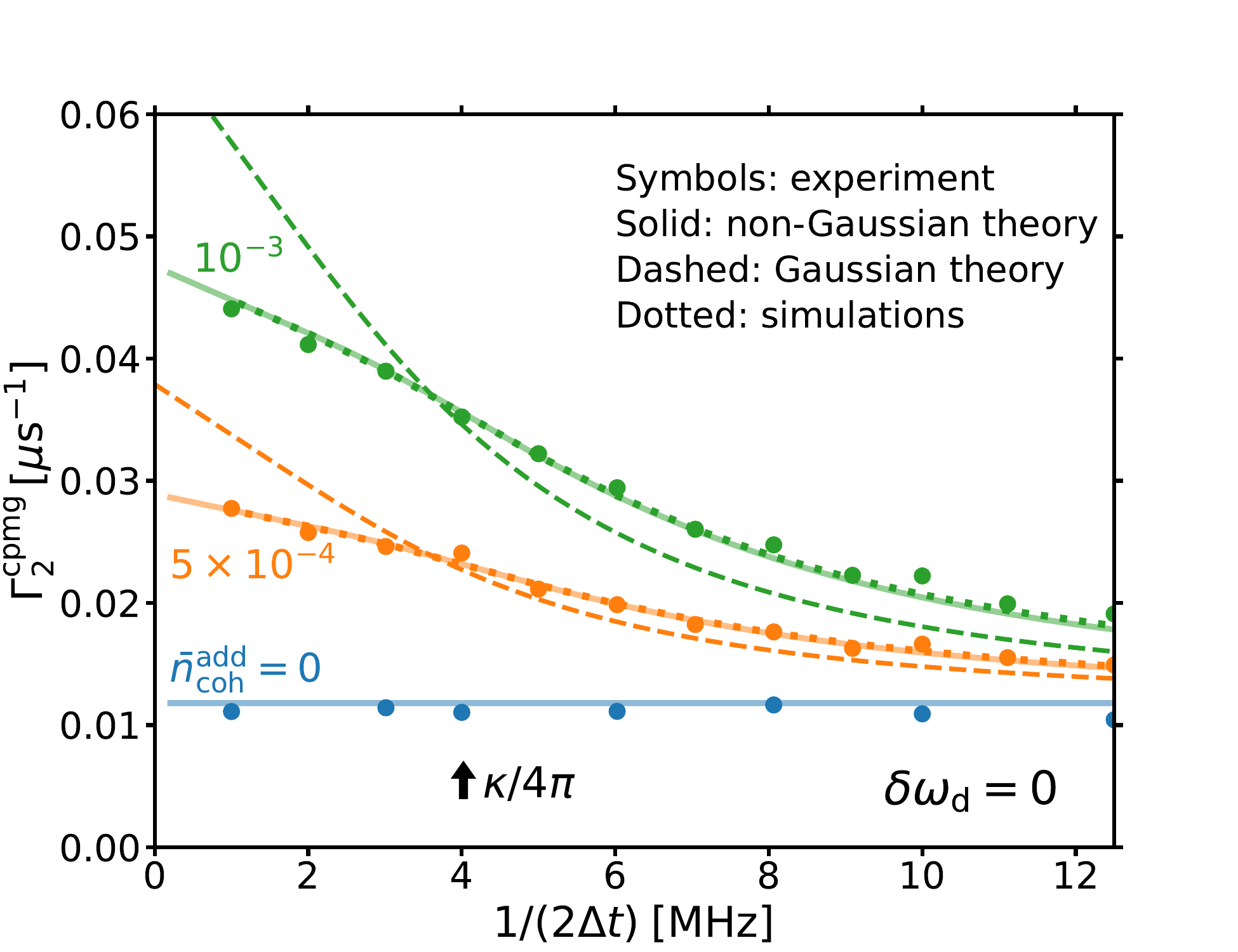}
    \caption{CPMG dephasing rate from shot noise of coherent photons. The   experimental data (symbols) correspond to added coherent photon populations of $\ncoh^{\rm add}=0$ (blue), $5\times10^{-4}$ (orange), and $10^{-3}$ (green). Solid lines depict our theory prediction while the dashed lines indicate the filter-function theory prediction. The dotted lines depict numerical results that include the shape and duration of experimental $\pi$-pulses.}
    \label{fig:coh_expt_results}
\end{figure}

 The symbols in Fig.~\ref{fig:coh_expt_results} show the measured CPMG dephasing rates $\Gamma_2^{\rm cpmg}$ for various interpulse periods $\Delta t$ (converted into the frequency $1/2\Delta t$) for added coherent photon populations  $\ncoh^{\rm add}=0$ (blue), $5\times10^{-4}$ (orange) and $10^{-3}$ (green) due to applied on-resonance drive, $\delta\omega_{\rm d}=0$. Note that blue symbols are different from  those in Fig.~\ref{fig:thermal_expt_results}(b) (a different experiment a few hours apart). 
We now fit the experimental data in Fig.~\ref{fig:coh_expt_results} as $\Gamma_\varphi^{\rm coh}(\Delta t, \ncoh) + \Delta \Gamma_2$ (solid lines), 
 where $\Gamma_\varphi^{\rm coh}$ is given by Eqs.~\eqref{coh-main-result-p1}--\eqref{coh-main-result-p2} and we have neglected the contribution due to thermal photons. Using four fitting parameters (three values of $\ncoh$ and a common $\Delta \Gamma_2$), we obtain $\ncoh = 2\times 10^{-8}$, $4.9\times10^{-4}$, and $1.03\times10^{-3}$, which are very close to the intended values of $\ncoh^{\rm add}$, and $\Delta \Gamma_2 = (84.8\,\mu{\rm s})^{-1}$. The difference in $\Delta \Gamma_2$ compared with fitting data in Fig.~\ref{fig:thermal_expt_results}(b) is due to  $\nth^{\rm amb}$. Note that if in Fig.~\ref{fig:coh_expt_results} we include $\nth^{\rm amb}$ as the fifth fitting parameter, we obtain $\nth^{\rm amb}\simeq 10^{-10}$, while if we fit only the blue symbols (with $\ncoh=0$), we obtain $\nth^{\rm amb} = (0.5\pm0.4)\times 10^{-4}$. These results are consistent with the method inaccuracy of about $10^{-4}$. 
 
In Fig.~\ref{fig:coh_expt_results} we again see a very good agreement between the experiment and our analytical theory (orange and green solid lines), which is in sharp contrast to the filter-function theory shown by the dashed lines (using the same $\ncoh$ and $\Delta \Gamma_2$). The dotted lines depict numerical results where we include the shape and duration (25~ns) of the experimental qubit pulses; numerics practically coincide with analytics.

{\it Conclusions.---}We have derived and validated formulas for the CPMG dephasing rate of a qubit due to thermal and coherent photons inside a resonator. These formulas are valid for an arbitrary ratio $2\chi/\kappa$, accounting for a non-Gaussian noise. We have also demonstrated that these formulas and CPMG data can be used to measure average thermal and coherent intracavity photon populations at the level of $\sim 10^{-4}$ photons.

We thank the Google Quantum AI team for fabrication of the measured device and for building and maintaining the hardware and software infrastructure used in this work. We also thank Mark I. Dykman for a significant contribution during the early stage of this work and for useful discussions.

\clearpage
\onecolumngrid
\begin{center}
\textbf{\large Supplemental Material for  "Measurement of small photon numbers in circuit QED resonators"}
\end{center}

\twocolumngrid

\newcommand{\cohformula}{3}
\newcommand{\cohdef}{1}
\newcommand{\whitenoisecorr}{6}
\newcommand{\EOMalpha}{4}
\newcommand{\thermalmainresultOne}{13}
\newcommand{\thermalmainresultTwo}{14}
\newcommand{\cohmainresultOne}{16}
\newcommand{\cohmainresultTwo}{17}
\newcommand{\Gammaphiapprox}{7}
\newcommand{\Ath}{8}
\newcommand{\gammasZeroOne}{5}
\newcommand{\alphacondcohcase}{15}
\newcommand{\Gammaphifilterfunc}{2}

\counterwithin*{equation}{section}

\setcounter{figure}{0}
\makeatletter 
\renewcommand{\thefigure}{S\@arabic\c@figure}
\renewcommand\theequation{\thesection\@arabic\c@equation}

\renewcommand\thesection{\Alph{section}}
\renewcommand\thesubsection{\thesection.\arabic{subsection}}

\renewcommand{\bibnumfmt}[1]{[S#1]} 
\renewcommand{\citenumfont}[1]{S#1}
\setcounter{page}{1} 

\section{Qubit coherence formula, E\MakeLowercase{q}.~(\cohformula)}
In this section we derive Eq.~(\cohformula) of the main text. We do this in two different ways. One approach is based on the Lindblad evolution equation of the resonator-qubit density matrix and the other approach is based on the wavefunction of the full system (qubit, resonator and the resonator bath characterized by coherent states propagating away from the cavity). We shall assume that the qubit decoherence is only due to photon shot noise and the qubit pulses are ideal and instantaneous.

\subsection{Lindblad approach}

 Our starting point is the Lindblad evolution equation for the off-diagonal block $\rho_{01}$ of the resonator-qubit density matrix,
\begin{align}
\label{SI:rho_01_eqn}
        \dot \rho_{01}(t) =&\, i\chi(\hat n \rho_{01} + \rho_{01} \hat n)  + i\delta\omega_{\rm d}[\hat{n},\rho_{01}] \nonumber \\
    &\, +\sqrt{\kappa}\,[F_{\rm d}(t)a^\dagger - F^*_{\rm d}(t)a, \rho_{01}] \nonumber \\
    &\,+\kappa(\bar n_{\rm th}+1)\left[a \rho_{01} a^\dagger - \frac{1}{2}a^\dagger a\rho_{01} - \frac{1}{2}\rho_{01} a^\dagger a\right]\nonumber \\
    &\,+\kappa\bar n_{\rm th}\left[a^\dagger \rho_{01} a - \frac{1}{2}a a^\dagger \rho_{01} - \frac{1}{2}\rho_{01} a a^\dagger\right],
\end{align}
where $a$ and $a^\dagger$ are the annihilation and creation harmonic oscillator (resonator) operators and $\hat{n} = a^\dagger a$. Equation~\eqref{SI:rho_01_eqn} describes the evolution of $\rho_{01}$ in between the $\pi$-pulses. At the moment $t_{\rm p}$ of a $\pi$-pulse, $\rho_{01}$ changes discontinuously,
\begin{align}
    \label{rho_discontinuous_cond}
    \rho_{01}(t_{\rm p}+0) =&\, \rho_{01}^\dagger(t_{\rm p}-0).
\end{align}

To analyze the qubit coherence, it is more convenient to consider a time-continuous operator $\tilde\rho_{01}(t)$ that coincides with $\rho_{01}(t)$ up to a Hermitian-conjugation operation. Specifically, $\tilde\rho_{01}(t) = \rho_{01}(t)$ for the times before the first $\pi$-pulse; $\tilde\rho_{01}(t) = [\rho_{01}(t)]^\dagger$ for the times in between the first and second $\pi$-pulses; $\tilde\rho_{01}(t) = \rho_{01}(t)$ for the times in between the second and third $\pi$-pulses; and so on. The evolution equation of $\tilde{\rho}_{01}(t)$ reads as 
\begin{align}
\label{SI:rho_01_eqn-v2}
        \dot{\tilde{\rho}}_{01}(t) =&\, i\tilde{\chi}(t)(\hat n \tilde\rho_{01} + \tilde\rho_{01} \hat n)  + i\delta\omega_{\rm d}[\hat{n},\tilde{\rho}_{01}] \nonumber \\
    &\, +\sqrt{\kappa}\,[F_{\rm d}(t)a^\dagger - F^*_{\rm d}(t)a, \tilde{\rho}_{01}] \nonumber \\
    &\,+ \kappa(\bar n_{\rm th}+1)\left[a \tilde\rho_{01} a^\dagger - \frac{1}{2}a^\dagger a\tilde\rho_{01} - \frac{1}{2}\tilde\rho_{01} a^\dagger a\right]\nonumber \\
    &\,+ \kappa\bar n_{\rm th}\left[a^\dagger \tilde\rho_{01} a - \frac{1}{2}a a^\dagger \tilde\rho_{01} - \frac{1}{2}\tilde\rho_{01} a a^\dagger\right],
\end{align}
where $\tilde\chi(t)=\pm\chi$ is, as mentioned in the main text, a piecewise constant  function that starts with  $\tilde\chi(0)=\chi$ and flips sign after each $\pi$-pulse. 

We point out that the magnitude of the qubit coherence is not affected if we use $\rho_{01}(t)$ or $\tilde\rho_{01}(t)$ because of the absolute value operation in Eq.~(\cohdef) of the main text. 

Equation~\eqref{SI:rho_01_eqn-v2} can be solved using the (generalized) positive P-representation defined as~\cite{S-GardinerBook, S-Gambetta2006}
\begin{align}
\label{SI:P-function-def}
    \tilde\rho_{01}(t) = \int {\rm d}^2\alpha{\rm d}^2\beta\, \frac{|\alpha\rangle\langle \beta^*|}{\langle \beta^*|\alpha \rangle} P_{01}(\alpha, \beta, t) .   
\end{align}
Here $P_{01}(\alpha,\beta,t) = \mathcal{P}[\tilde{\rho}_{01}(t)]$ is the positive P-representation function of $\tilde\rho_{01}(t)$ with independent complex-valued variables $\alpha$ and $\beta$, and ${\rm d}^2\alpha{\rm d}^2\beta = {\rm d}({\rm Re}\, \alpha){\rm d}({\rm Im}\, \alpha) {\rm d}({\rm Re}\, \beta){\rm d}({\rm Im}\, \beta)$.  $\mathcal{P}[\cdot]$ denotes the mapping from $\tilde\rho_{01}$ to its positive P-representation function, $P_{01}$. Applying the $\mathcal{P}$ transformation to Eq.~\eqref{SI:rho_01_eqn-v2} and using the relations
\begin{align}
    \mathcal{P}[a\tilde\rho_{01}] &= \alpha P_{01}, \;\; \mathcal{P}[\tilde\rho_{01} a] = (\alpha - \partial_{\beta})P_{01}, \nonumber \\
    \mathcal{P}[\tilde\rho_{01} a^\dagger] &= \beta P_{01},\;\; \mathcal{P}[a^\dagger \tilde\rho_{01}] = (\beta - \partial_{\alpha})P_{01},
    \end{align}
we obtain the following evolution equation for $P_{01}$: %
\begin{align}
\label{SI:P01-eqn}
    \partial_t P_{01}(&\alpha, \beta, t) = 2i\tilde\chi(t) \alpha\beta P_{01}\nonumber \\ 
    &\; - \partial_\alpha\Big[\Big(-\big[\kappa/2-i\tilde\chi(t)-i\delta\omega_{\rm d}\big]\alpha + \sqrt{\kappa}F_{\rm d}(t)\Big)P_{01}\Big]   \nonumber \\ 
    &\; - \partial_\beta\Big[\Big(-\big[\kappa/2-i\tilde\chi(t)+i\delta\omega_{\rm d}\big]\beta + \sqrt{\kappa}F_{\rm d}^*(t)\Big)P_{01}\Big]   \nonumber \\ 
    &\; + \kappa \bar n_{\rm th} \partial_{\alpha \beta}^2 P_{01}. 
\end{align}
%
The solution of Eq.~\eqref{SI:P01-eqn} can be written as 
\begin{align}
\label{SI:P01-sol}
    P_{01}(\alpha, \beta, t) =&\;   \int {\rm d}^2\alpha_0{\rm d}^2\beta_0\, \bigg\langle\exp\left[\int_0^{t} {\rm d}t'\, 2i\tilde\chi(t)\alpha(t')\beta(t')\right]   \nonumber\\
    &\; \times \delta^2(\alpha - \alpha(t))\, \delta^2(\beta - \beta(t)) \bigg\rangle P_{01}(\alpha_0,\beta_0,0),
\end{align}
where $\delta^2(z)$ is short notation for $\delta^2(z) = \delta({\rm Re}\, z)\delta({\rm Im}\, z)$, and $\alpha(t)$ and $\beta(t)$ are described by 
\begin{align}
\label{SI:alpha-beta-eqns}
    \dot \alpha(t) = -[\kappa/2 - i\tilde\chi(t)-i\delta\omega_{\rm d}]\alpha + \sqrt{\kappa}\, \xi_{\rm th}(t) + \sqrt{\kappa}\,F_{\rm d}(t) \, , \nonumber \\
    \dot \beta(t) = -[\kappa/2 - i\tilde\chi(t)+i\delta\omega_{\rm d}]\beta + \sqrt{\kappa}\,  \xi_{\rm th}^*(t) + \sqrt{\kappa}\,F^*_{\rm d}(t),
\end{align}
with the initial conditions $\alpha(0)=\alpha_0$ and $\beta(0)=\beta_0$. The operation $\langle\cdot \rangle$ in Eq.~\eqref{SI:P01-sol} indicates averaging over the complex-valued noise $\xi_{\rm th}(t)$ that is defined by the time-correlators of Eq.~(\whitenoisecorr) of the main text. 

In terms of the positive P-representation function $P_{01}$, the coherence is given by    
\begin{align}
\label{SI:coh_in_P_representation}
    \mathcal{C}(t_{\rm cpmg}) = \left|\int {\rm d}^2\alpha{\rm d}^2\beta\, 2P_{01}(\alpha, \beta, t_{\rm cpmg})\right|. 
\end{align}
To evaluate the coherence, we insert the solution Eq.~\eqref{SI:P01-sol} for $P_{01}(\alpha, \beta, t)$ into Eq.~\eqref{SI:coh_in_P_representation} and perform the integrals over $\alpha$ and $\beta$ (these integrations cancel out the delta-functions of Eq.~\eqref{SI:P01-sol}) and obtain 
\begin{align}
\label{SI:coh_in_P_representation-p2}
    \mathcal{C}(&t_{\rm cpmg}) = \nonumber \\
    &\bigg|\int {\rm d}^2\alpha_0{\rm d}^2\beta_0\, \bigg\langle\exp\left[\int_0^{t_{\rm cpmg}} {\rm d}t\, 2i\tilde\chi(t)\alpha(t)\beta(t)\right] \bigg\rangle \nonumber \\
    &\times 2P_{01}(\alpha_0,\beta_0,0)\bigg|,
\end{align}

Assuming that the resonator is initially in a coherent state $|\alpha_{\rm init}\rangle \langle\alpha_{\rm init}|$; that is,  $P_{01}(\alpha_0,\beta_0,0)=\delta^2(\alpha_0-\alpha_{\rm init})\delta^2(\beta_0-\alpha_{\rm init }^*)/2$, we can easily perform the integrals over $\alpha_0$  and $\beta_0$ in Eq.~\eqref{SI:coh_in_P_representation-p2} and obtain
\begin{align}
\label{SI:eqn3_derivation_final_result}
    \mathcal{C}(t_{\rm cpmg}) = \bigg|\bigg\langle\exp\left[\int_0^{t_{\rm cpmg}} {\rm d}t\, 2i\tilde\chi(t)\alpha(t)\beta(t)\right] \bigg\rangle\bigg|,
\end{align}
where the trajectories $\alpha(t)$ and $\beta(t)$ start at $\alpha(0)=\alpha_{\rm init}$ and $\beta(0)=\alpha_{\rm init}^*$. 
Equation~\eqref{SI:eqn3_derivation_final_result} coincides with Eq.~(\cohformula) of the main text if we replace $\alpha(t)$ and $\beta(t)$ by $\alpha_0(t)$ and $\alpha_1^*(t)$, respectively.

\subsection{Wavefunction overlap method}
\label{wave_function_overlap_method}

In this approach~\cite{S-Korotkov2016} the qubit dephasing comes from the overlap of two wavefunctions $|\alpha_0(t)\rangle|\psi^{|0\rangle}_{\rm bath}(t)\rangle$ and $|\alpha_1(t)\rangle|\psi^{|1\rangle}_{\rm bath}(t)\rangle$, where $|\alpha_q(t)\rangle$ and $|\psi^{|q\rangle}_{\rm bath}(t)\rangle$ respectively describe the coherent resonator state and the state of the resonator bath (which in our case is as an infinite transmission line that is weakly coupled to the resonator) when the qubit is in the state $|q\rangle$ ($q=0,1$). The resonator bath state,
\begin{align}
\label{SI:psi-env-0}
    |\psi_{\rm bath}^{|q\rangle}(t) \rangle =&\, \big|\sqrt{\kappa\delta t}\, \alpha_q(t)\big\rangle \, \big|\sqrt{\kappa\delta t}\, \alpha_q(t-\delta t)\big\rangle \, \nonumber \\
    &\,\times \big|\sqrt{\kappa\delta t}\, \alpha_q(t-2\delta t)\big\rangle \, \cdots \big|\sqrt{\kappa\delta t}\, \alpha_q(0)\big\rangle,
\end{align}
can be interpreted as the history of the intracavity state that is imprinted into the resonator-bath  state in the form of coherent states propagating away from the resonator. Each piece of the flying away ``history tail'', e.g. $|\sqrt{\kappa\delta t}\,\alpha_q(t)\rangle$, is due to the intracavity state $|\alpha_q(t)\rangle$ leaking out through a low-transparency  mirror (a beam splitter with transmission amplitude $\sqrt{\kappa\delta t}$ where $\delta t\ll 1/\kappa$)  at the time $t$.  

Since the dispersive interaction, $H_{\rm int}=-\chi \sigma_z \hat{n}$, does not induce evolution of the qubit basis states (i.e., if the qubit is initially in state $|0\rangle$, it remains in the same state with some accumulated phase factor) and the $\pi$-pulses of the CPMG sequence flip the qubit states ($|0\rangle \leftrightarrow |1\rangle$), we can pretend that the qubit is always in the state $|0\rangle$ or $|1\rangle$ while the effect of the $\pi$-pulses can be accounted for by making the parameter $\chi$ time-dependent as discussed in the main text; i.e., $\chi\to\tilde\chi(t)=\pm\chi$. In this approximation, the wavefunction of the full system is given by ($c_0=c_1=1/\sqrt{2}$ for the CPMG sequence)
\begin{align}
\label{SI:full_wavefunction}
    |\Psi(t)\rangle =&\,  c_0 e^{-i\varphi_0(t)}|0\rangle|\alpha_0(t)\rangle |\psi_{\rm bath}^{|0\rangle}\rangle \nonumber \\
    &\, + c_1 e^{-i\varphi_1(t)}|1\rangle|\alpha_1(t)\rangle |\psi_{\rm bath}^{|1\rangle}\rangle,
\end{align}
where the equations of motion for $\alpha_0(t)$ and $\alpha_1(t)$ are given in Eq.~(\EOMalpha) of the main text and $\varphi_q(t)$ is a phase angle coming from the resonator evolution (see Appendix A of Ref.~\cite{S-Korotkov2016}), 
\begin{align}
\label{SI:phase_coherent_evol}
    \dot{\varphi}_q(t) = {\rm Re}(\varepsilon^*\alpha_q), \;\; \varepsilon(t) =  i\sqrt{\kappa}\,[\xi_{\rm th}(t) + F_{\rm d}(t)].  
\end{align}
We point out that Eqs.~\eqref{SI:full_wavefunction}--\eqref{SI:phase_coherent_evol} hold for a given realization of the noisy resonator drive $\xi_{\rm th}(t)$. Next, we use Eq.~(59) of Ref.~\cite{S-Korotkov2016} to write down the qubit off-diagonal density matrix element $\rho^{\rm q}_{10}(t_{\rm cpmg})$,
\begin{align}
\label{SI:rho_10_qubit}
\rho^{\rm q}_{10}(t_{\rm cpmg}) = \left\langle\frac{1}{2}\exp\left[-\int_0^{t_{\rm cpmg}}{\rm d}t\, 2i\tilde\chi(t)\alpha_0^*(t)\alpha_1(t)\right]\right\rangle, 
\end{align}
which is obtained after tracing out the resonator and resonator-bath degrees of freedom from Eq.~\eqref{SI:full_wavefunction}. In Eq.~\eqref{SI:rho_10_qubit}, $\langle\cdot\rangle$ indicates averaging over the different noise realizations.  Finally, the visibility of the qubit coherence is given by $|2\rho_{10}^{\rm q}(t_{\rm cpmg})|$, which is the same as the expression Eq.~(\cohformula) of the main text.

Note that the main contribution to the decay of $\rho_{10}^{\rm q}(t_{\rm cpmg})$ is due to tracing over the ``history tail'', which gives $\exp [-\int_0^{t_{\rm cpmg}} (\kappa/2) |\alpha_0(t)-\alpha_1(t)|^2 dt]$, while the exponent in Eq.~\eqref{SI:rho_10_qubit} looks different because it also takes into account the evolution of $\langle \alpha_0(t_{\rm cpmg}) |\alpha_1 (t_{\rm cpmg})\rangle$ and phase factors \cite{S-Korotkov2016}. Derivation of the CPMG dephasing rates [Eqs.\ (\thermalmainresultOne)--(\thermalmainresultTwo) and (\cohmainresultOne)--(\cohmainresultTwo) of the main text] can also be done starting with this formula instead of Eq.~(\cohformula).

\section{CPMG dephasing rate from intracavity thermal photons: Lindblad approach}
\label{Lindblad_thermal}

Equations~(\thermalmainresultOne)--(\thermalmainresultTwo) of the main text provide the thermal-photon-induced dephasing rate in the limit $\nth\to 0$. For moderate values of $\nth$, we can still use Eq.~(\cohformula) to obtain the dephasing rate; however, in this section, we instead follow the approach based on the Lindblad master equation~\eqref{SI:rho_01_eqn-v2}. 

To solve the Lindblad master equation~\eqref{SI:rho_01_eqn-v2}, we use the Wigner transformation method~\cite{S-GardinerBook}, following the formalism developed in Appendix A of Ref.~\cite{Atalaya2019}. The Wigner function of $\tilde\rho_{01}$ can be defined as  
\begin{align}
    \mathcal{W}[\tilde\rho_{01}&(t)] = W_{01}(\alpha, \alpha^*, t) \nonumber \\
    =&\; \int  \frac{{\rm d}^2z}{\pi^2} \, {\rm Tr}[\tilde\rho_{01}(t)\exp(za^\dagger - z^*a)]e^{z^*\alpha-z\alpha^*},
\end{align}
where ${\rm d}^2z = d({\rm Re}\, z) d({\rm Im}\,z )$. Using the relations  
\begin{align}
    &\mathcal{W}[a\tilde\rho_{01}] = (\alpha + \frac{1}{2}\partial_{\alpha^*} )W_{01},\,\,\,   \mathcal{W}[\tilde\rho_{01}a^\dagger] = (\alpha^* + \frac{1}{2}\partial_{\alpha})W_{01}, 
    \nonumber \\
    &\mathcal{W}[a^\dagger\tilde\rho_{01}] = (\alpha^* - \frac{1}{2}\partial_{\alpha})W_{01}, \,\,\, \mathcal{W}[\tilde\rho_{01}a] = (\alpha - \frac{1}{2}\partial_{\alpha^*})W_{01}, \label{SI:Wigner_relations}
\end{align}
we apply the Wigner transformation to Eq.~\eqref{SI:rho_01_eqn-v2} and obtain the equation of motion for $W_{01}$
\begin{align}
\label{SI:W01_EOM}
    &\partial_t W_{01}(\alpha, \alpha^*, t) = i\tilde\chi(t)\left( (2|\alpha|^2-1)W_{01} - \frac{1}{2}\partial_{\alpha\alpha^*}^2 W_{01} \right) \nonumber \\
    &\; +\frac{\kappa}{2}\left[\partial_\alpha(\alpha W_{01}) + \partial_{\alpha^*}(\alpha^* W_{01}) + (2\bar n_{\rm th}+1)\partial_{\alpha\alpha^*}^2 W_{01}\right],
    \end{align}
where $W_{01}(\alpha, \alpha^*,t)$ is the Wigner function representation of $\tilde\rho_{01}(t)$ and $\alpha$ is a complex-valued independent variable. The qubit coherence is now given by 
\begin{align}
\label{SI:Coherence_def_Wigner}
    \mathcal{C}(t_{\rm cpmg}) = \left|\int {\rm d}^2\alpha \, 2W_{01}(\alpha, \alpha^*,t_{\rm cpmg})\right|.
\end{align}
Somewhat surprisingly, Eq.~\eqref{SI:W01_EOM} can be solved by using the following Gaussian {\it ansatz}:   
\begin{align}
\label{SI:Ansatz}
    W_{01}(\alpha, \alpha^*,t) = \frac{\exp\left[-A(t)-\frac{|\alpha|^2}{V(t)}\right]}{V(t)},
\end{align}
where $A(t)$ and $V(t)$ are complex-valued temporal functions. After inserting Eq.~\eqref{SI:Ansatz} into Eq.~\eqref{SI:W01_EOM}, we obtain the evolution equations for $A(t)$ and $V(t)$,
\begin{align}
    \dot A(t) =&\, -2i\tilde\chi(t)\,V + i\tilde\chi(t), \label{A_eq}\\
    \dot V(t) =&\, -\kappa V + 2i\tilde\chi(t)\, V^2 + \frac{\kappa(2\nth+1) -i\tilde\chi(t)}{2}.\label{V_eq}
\end{align}
The solution of Eq.~\eqref{V_eq} reads as (assuming $\tilde\chi(t)=\chi$)
\begin{align}
\label{SI:Vt_sol}
    V(t) = \frac{i\vartheta}{2\chi}\,\frac{\tanh[(t-t_{\rm p})\vartheta]+\zeta}{1 + \zeta\tanh[(t-t_{\rm p})\vartheta]}
    -\frac{\kappa i}{4\chi}, 
\end{align}
where $t\in(t_{\rm p}, t_{\rm p}+\Delta t)$, $t_{\rm p}$ indicates the moment of one of the qubit $\pi$-pulses, and $\vartheta$ and $\zeta$ are parameters that depend on the thermal photon population parameter $\nth$, 
\begin{align}
\label{SI:vartheta}
    \vartheta =&\;\frac{1}{2}\sqrt{\kappa^2 - 4\chi^2 - 4i\chi\kappa(2\nth+1)}\;\;\; ({\rm Re}\, \vartheta>0),\\
\zeta =&\; \frac{\kappa - 4i\chi V(t_{\rm p})}{2\vartheta}, \label{SI:z_def}
\end{align}
and $V(t_{\rm p})$ also implicitly depends on $\nth$. $V(t_{\rm p})$ is so far a free parameter. It will be found using the quasi-steady-state condition $V(t_{\rm p}+\Delta t) = [V(t_{\rm p})]^*$, as discussed below. 

From Eq.~\eqref{A_eq}, we obtain the real part of the average rate of change of $A(t)$, 
\begin{align}
\label{SI:A_rel_change}
    {\rm Re}\left[\frac{\Delta A}{\Delta t}\right] = -{\rm Re} \Bigg[\frac{\int_{t_{\rm p}}^{t_{\rm p}+\Delta t} {\rm d}t\, 2i\tilde\chi(t)V(t)}{\Delta t}\Bigg]. 
\end{align}
From Eqs.~\eqref{SI:Coherence_def_Wigner} and~\eqref{SI:Ansatz}, it is easy to see that the photon-induced dephasing rate $\Gamma_\varphi^{\rm th}$ is given by Eq.~\eqref{SI:A_rel_change} if $t_{\rm p}$ is in the quasi-steady-state regime of the qubit coherence. In such a case, using the above solution for $V(t)$ [Eq.~\eqref{SI:Vt_sol}] to evaluate the integral in Eq.~\eqref{SI:A_rel_change} with $\tilde\chi(t)=\chi$, we obtain    
\begin{align}
\label{SI:Gamma_varphi_sol_v0}
    \Gamma_\varphi^{\rm th} = \frac{{\rm Re}\,\ln\left[\cosh(\vartheta\Delta t) + \zeta\, \sinh(\vartheta\Delta t)\right]}{\Delta t} - \frac{\kappa}{2}.
\end{align}

Let us consider first the low-frequency limit of the CPMG dephasing rate. Taking the limit $\Delta t\to \infty$ of Eq.~\eqref{SI:Gamma_varphi_sol_v0}, we obtain 
\begin{align}
\label{SI:Gamma_Ramsey_th}
    \Gamma_\varphi^{\rm th}(\Delta t\to \infty, \nth) = \frac{\kappa}{2}\, \R\left[\sqrt{\left(1 - \frac{2i\chi}{\kappa} \right)^2 - \frac{8i\chi}{\kappa}\,\bar{n}_{\rm th}}\, - 1\right],
\end{align}
which agrees with Eqs.~(11)--(12) of Ref.~\cite{S-Dykman1987} (where $Q(t)\sim {\rm Tr}[\rho_{01}(t)]$,  $\Gamma_\varkappa=\kappa/2$, $V_{\varkappa}=2\chi$ and $\bar{n}_{\varkappa}=\nth$) and with Eq.~(44) of Ref.~\cite{S-Clerk2007}. Furthermore, if we consider the small-photon limit, $\nth\ll1$, we find that Eq.~\eqref{SI:Gamma_Ramsey_th} leads to the prefactor of Eq.~(\thermalmainresultOne) of the main text. 

For an arbitrary interpulse period $\Delta t$, the photon-induced dephasing rate is given by Eq.~\eqref{SI:Gamma_varphi_sol_v0}. To evaluate it, we first need to find $V(t_{\rm p})$ that enters in the definition of the parameter $\zeta$, see Eq.~\eqref{SI:z_def}. We notice that if $V(t)$ satisfies the condition $V(t+\Delta t) =[V(t)]^*$, then the value of $\Gamma_\varphi^{\rm th} = \R[\Delta A/\Delta t]$ given by Eq.~\eqref{SI:A_rel_change} is independent of $t_{\rm p}$, which is  necessary for an exponential decay of the coherence (quasi-steady-state regime). From this condition, we obtain the following nonlinear algebraic equation for $V(t_{\rm p})$: 
\begin{align}
\label{SI:V0_eqn}
        [V(t_{\rm p})]^* =&\, V(t_{\rm p}+\Delta t) \nonumber \\
        =&\, \frac{i\vartheta}{2\chi} \, \frac{\tanh(\vartheta\Delta t) + \frac{\kappa - 4i\chi V(t_{\rm p})}{2\vartheta}}{1 + \frac{\kappa - 4i\chi V(t_{\rm p})}{2\vartheta} \, \tanh(\vartheta\Delta t)}-\frac{\kappa i}{4\chi}.
\end{align}

\begin{figure}[t!]
    \centering
    \includegraphics[width=0.85\columnwidth,trim=1cm 0.5cm 1cm 0cm,clip=True]{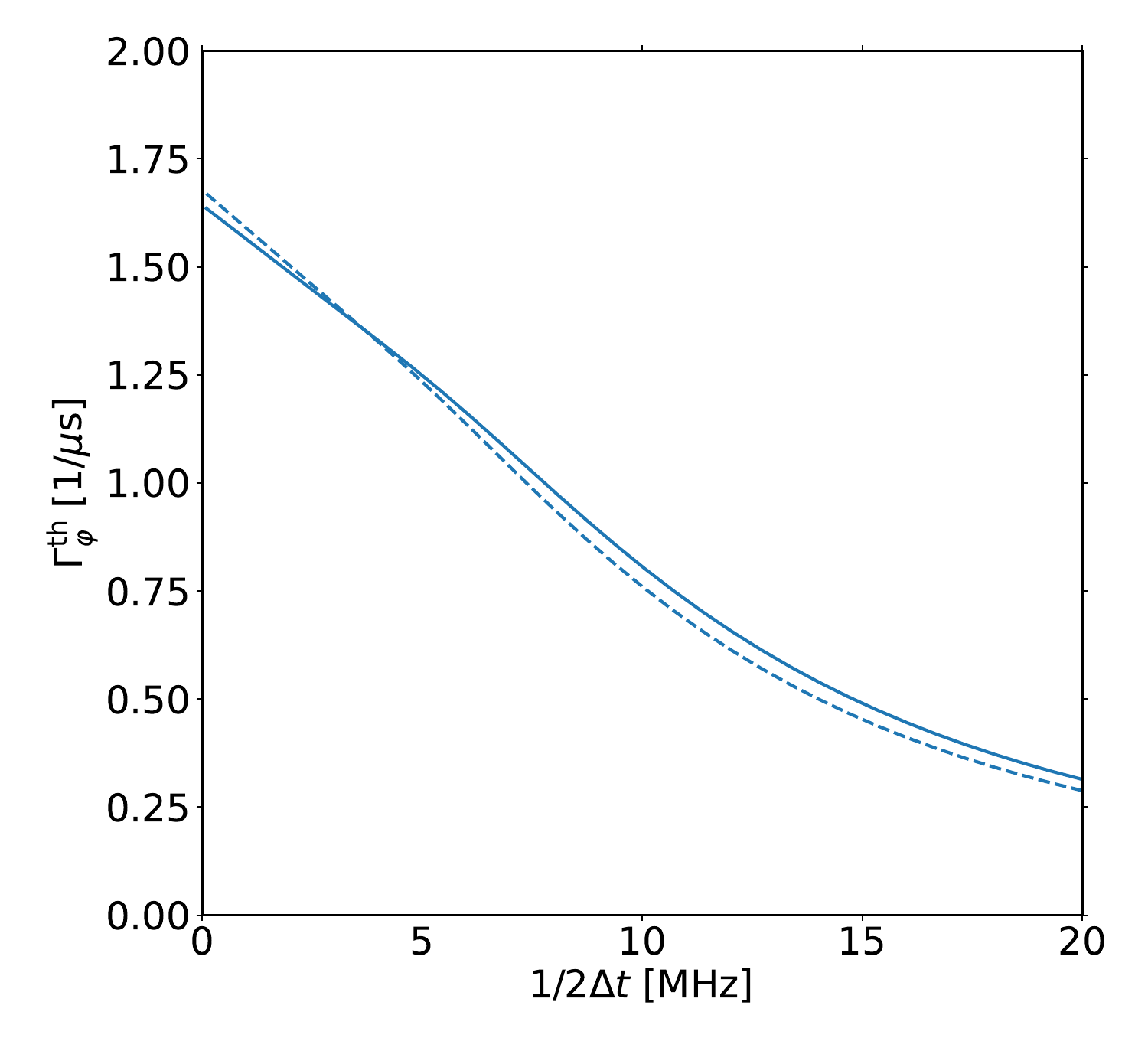}
    \caption{Photon-induced dephasing rate for $\nth=0.1$. The solid line depicts the theory prediction Eq.~\eqref{SI:Gamma_varphi_sol_v0} with $\vartheta$ given by Eq.~\eqref{SI:vartheta} and $\zeta$ given by Eq.~\eqref{SI:z_def} with $V(t_{\rm p})$ obtained from the solution of Eq.~\eqref{SI:V0_eqn}. Dashed line depicts the small-thermal-population limit formula for $\Gamma_\varphi^{\rm th}$   [Eqs.~(\thermalmainresultOne)--(\thermalmainresultTwo) of the main text]. We use $\kappa=(19\,{\rm ns})^{-1}$ and $2\chi=2\pi\times5.7\,{\rm MHz}$.}
    \label{fig:Gamma_phi_vs_fcpmg_0.1_nth}
\end{figure}

The photon-induced dephasing rate $\Gamma^{\rm th}_\varphi$ for moderate values of  the thermal photon number $\bar n_{\rm th}$ can be obtained from Eq.~\eqref{SI:Gamma_varphi_sol_v0} with $\vartheta$ given by Eq.~\eqref{SI:vartheta} and $\zeta$ given by Eq.~\eqref{SI:z_def}, where $V(t_{\rm p})$ is obtained from the solution of Eq.~\eqref{SI:V0_eqn}. As shown by Fig.~\ref{fig:Gamma_phi_vs_fcpmg_0.1_nth}, we find that the $\Gamma_\varphi^{\rm th}$ formula given in the main text [Eqs.~(\thermalmainresultOne)--(\thermalmainresultTwo)] is still a good approximation for $\nth=0.1$ and experimental values of $2\chi$ and $\kappa$.

\newpage 
\vspace{1cm}
\section{CPMG dephasing rate for arbitrary drive frequency detuning ($\delta \omega_{\rm d}\neq 0$)}

The dephasing rate formula for vanishing detuning ($\delta\omega_{\rm d}=0$) is given in the main text by Eq.~(\cohmainresultOne)--(\cohmainresultTwo). For an arbitrary detuning, $\delta\omega_{\rm d}\neq0$, the dephasing rate reads as  
\begin{align}
\label{SI:coh-result-arbitrary_dw-p1}
    \Gamma_\varphi^{\rm coh}(\Delta t, \delta \omega_{\rm d})  =&\;\frac{8\chi^2\,\bar n_{|0\rangle}\bar n_{|1\rangle}}{\kappa\,\bar n_{\max}} \, \mathcal{R}_{\rm coh}(\Delta t, \delta \omega_{\rm d}), \\
        \bar n_{|q\rangle} =&\; \frac{\kappa |F^{\rm dc}_{\rm d}|^2}{(\kappa/2)^2 + [\delta \omega_{\rm d} + (-1)^q\chi]^2}, 
\end{align}
where $\bar n_{\rm max}\equiv \kappa |F_{\rm d}^{\rm dc}|^2/(\kappa/2)^2$ is the maximum photon number population for a given drive power (without CPMG). The second factor of Eq.~\eqref{SI:coh-result-arbitrary_dw-p1} reads as 
\begin{widetext}
\begin{align}
\label{SI:coh-result-arbitrary_dw-p2}
    \mathcal{R}_{\rm coh}(\Delta t, \delta \omega_{\rm d}) = 1 - \frac{\mathbf{\mathcal{B}} - 4(\kappa\Delta t)^{-1} [\cosh(\kappa\Delta t/2) - \cos(\chi\Delta t)\cos(\delta \omega_{\rm d}\Delta t)]\sinh(\kappa\Delta t/2)e^{-\kappa\Delta t}}{(1+e^{-2\kappa\Delta t})/2 - \cos(2\delta\omega_{\rm d}\Delta t)e^{-\kappa\Delta t}},
\end{align}
\end{widetext}
where $\mathcal{B}$ is given by 
%
\begin{align}
\label{SI:coh-result-arbitrary_dw-p3}
\mathbf{\mathcal{B}} &=\R\bigg[\frac{(1-e^{-\gamma_0\Delta t})(1-e^{-\gamma_1\Delta t})(1-e^{-(\gamma_0^*+\gamma_1^*)\Delta t})}{\gamma_1\Delta t} \nonumber \\
&\;+ \frac{(1-e^{-\gamma_0^*\Delta t})(1-e^{-\gamma_1^*\Delta t})(1-e^{-(\gamma_0+\gamma_1)\Delta t})}{\gamma_0^*\Delta t}  \nonumber \\
&\;- \frac{(1-e^{-\gamma_0\Delta t})(1-e^{-\gamma_1^*\Delta t})(1-e^{-(\gamma_0^*+\gamma_1)\Delta t})}{(\gamma_0^*+\gamma_1)\Delta t}\bigg]
\end{align}
and $\gamma_q = \kappa/2 -i[\delta\omega_{\rm d} + (-1)^q\chi]$. 
Note that in the low-frequency limit, $\Delta t \to \infty$, we get $\mathcal{R}_{\rm coh}=1$ and the CPMG dephasing rate (\ref{SI:coh-result-arbitrary_dw-p1}) agrees with Eq.~(69) of Ref.~\cite{S-Clerk2007}, Eq.~(5.20) of Ref.~\cite{S-Gambetta2006}, and Eq.~(17) of Ref.~\cite{S-Korotkov2015}.

To obtain the above dephasing rate formula \eqref{SI:coh-result-arbitrary_dw-p1}--\eqref{SI:coh-result-arbitrary_dw-p3}, we use Eqs.~(\Gammaphiapprox)--(\Ath) of the main text with $\tilde\chi(t)=\chi$ and with the trajectories $\alpha_0(t)$ and $\alpha_1(t)$ given by the solutions of Eqs.~(\EOMalpha)--(\gammasZeroOne) (where the noise terms are dropped out): 
\begin{align}
\label{SI:alpha0_1_sol_coh_case}
    \alpha_q(t) = e^{-\gamma_q(t-t_{\rm p})}\alpha_{ q}(t_{\rm p}) + \sqrt{\kappa}\,F^{\rm dc}_{\rm d}\,\frac{1-e^{-\gamma_q(t-t_{\rm p})}}{\gamma_q},
\end{align}
where $t\in(t_{\rm p}, t_{\rm p}+\Delta t)$ and we have assumed that the coherent drive field is constant, $F_{\rm d}(t)=F_{\rm d}^{\rm dc}$. To find the initial conditions $\alpha_0(t_{\rm p})$ and $\alpha_1(t_{\rm p})$ in Eq.~\eqref{SI:alpha0_1_sol_coh_case}, we use  Eq.~(\alphacondcohcase) of the main text with $t=t_{\rm p}$ that lead to a linear system of equations for $\alpha_0(t_{\rm p})$ and $\alpha_1(t_{\rm p})$. After finding the initial conditions $\alpha_{q}(t_{\rm p})$, we use  Eq.~\eqref{SI:alpha0_1_sol_coh_case} to evaluate the integral in Eq.~(\Gammaphiapprox) of the main text, thus obtaining Eqs.~\eqref{SI:coh-result-arbitrary_dw-p1}--\eqref{SI:coh-result-arbitrary_dw-p3}.

\section{Exponential decay regime of the qubit coherence}

\begin{figure*}
    \centering
    \includegraphics[width=0.85\columnwidth,trim=0cm 1cm 0cm 0cm,clip=True]{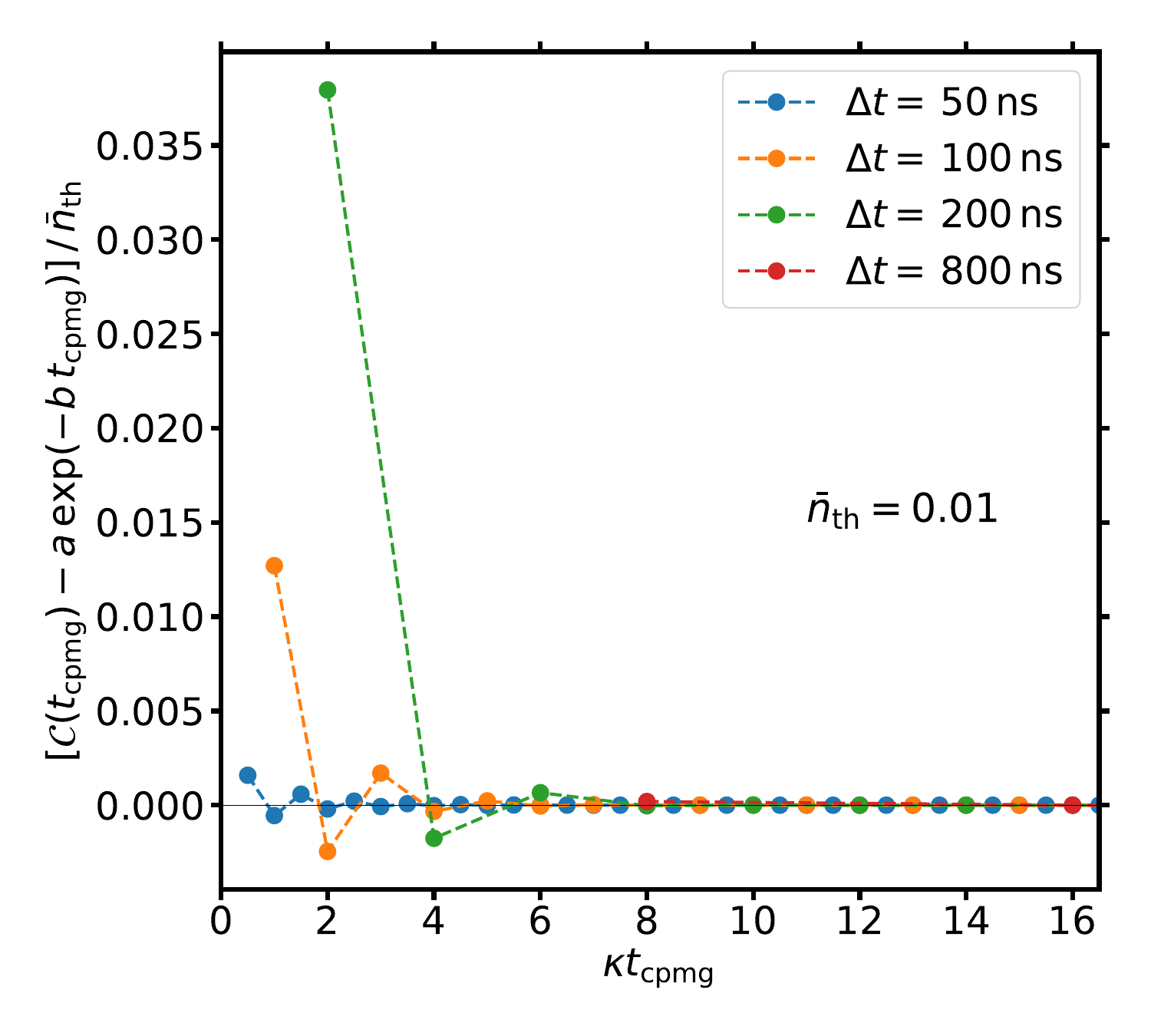}
    \includegraphics[width=0.85\columnwidth,trim=0cm 1cm 0cm 0cm,clip=True]{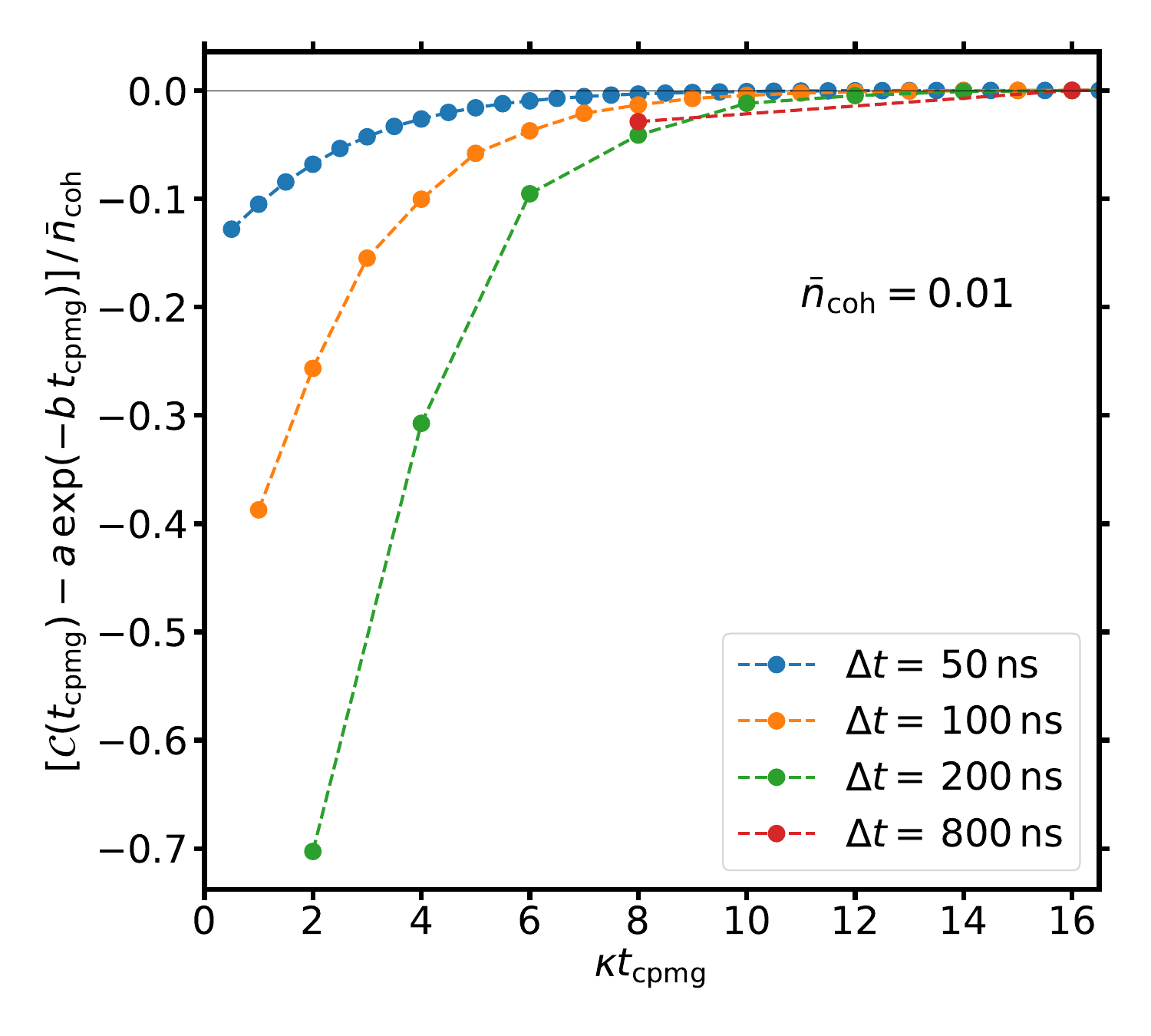}
    \caption{Qubit coherence {\it vs.} CPMG sequence duration (simulation). Left (Right) panel shows the coherence for the thermal (coherent) cases, where the first, second, third, {\it etc}. data points (from left to right) correspond to CPMG sequences with $N=1,2,3,$ {\it etc.}  $\pi$-pulses. We have subtracted from the coherence, $\mathcal{C}(t_{\rm cpmg})$, the (fitted) exponential decay $a\exp(-bt_{\rm cpmg})$ that holds at sufficiently large $N$, where $a,b$ are fitting parameters  with $a\lesssim1$ and $b$ agreeing well with the expected value from our analytical formulas for $\Gamma_\varphi$. These plots suggest that the exponential decay regime of the coherence occurs for $t_{\rm cpmg}\gtrsim 3\kappa^{-1}$ for the thermal case and for $t_{\rm cpmg}\gtrsim 6\kappa^{-1}$ for the coherent case. We use $\bar n_{\rm th}=\bar n_{\rm coh}=0.01$, $\delta\omega_{\rm d}=0$ (relevant for the coherent case), $2\chi/\kappa=1.5$ and $\kappa^{-1}=100\,{\rm ns}$. Similar results are obtained for other values of $2\chi/\kappa$; we point out that the largest deviation between $\mathcal{C}(t_{\rm cpmg})$ and the exponential fit decreases as we decrease $2\chi/\kappa$.}
    \label{fig:qubit_coherence_decay}
\end{figure*}

Figure~\ref{fig:qubit_coherence_decay} shows the qubit coherence after subtracting the asymptotic exponential decay dependence that holds at sufficiently long CPMG sequences, as a function of the CPMG sequence duration, $t_{\rm cpmg}$, for the cases of thermal (left panel) and coherent (right panel) photons in the resonator. We find that the exponential decay regime (quasi-steady-state regime) of the qubit coherence occurs roughly for $t_{\rm cpmg}\gtrsim 3/\kappa$ in the thermal case and $t_{\rm cpmg}\gtrsim 6/\kappa$ in the coherent case.

\section{Calibration of added intracavity photon population}

The added photon population (${\nth}^{\rm add}$ or ${\ncoh}^{\rm add}$) in the resonator is proportional to the power $\mathcal{P}_{\rm drive}$ of the resonator drive tone. The proportionality coefficient has to be determined in order to increase the intracavity photon population by a target amount. This calibration procedure is done using CPMG dephasing experiments with just one $\pi$-pulse ($N=1$) and with sequence durations $\Delta t$ that are much larger than the resonator decay rate (low-frequency limit: $1/(2\Delta t)\ll\kappa$). We use periods $\Delta t$ in the range from $1\,\mu{\rm s}$ to $40\,\mu{\rm s}$ (while $\kappa^{-1}=19.4\,{\rm ns}$) and then extract dephasing rate from fitting the coherence by the combination of exponential and Gaussian decays. These CPMG experiments are essentially the spin-echo experiments, with the unimportant difference that in the spin-echo experiments all qubit pulses are traditionally applied along the same Bloch-sphere direction.

\begin{figure*}
    \centering
    \includegraphics[width=0.85\columnwidth,trim=0cm 0cm 0cm 0cm,clip=True]{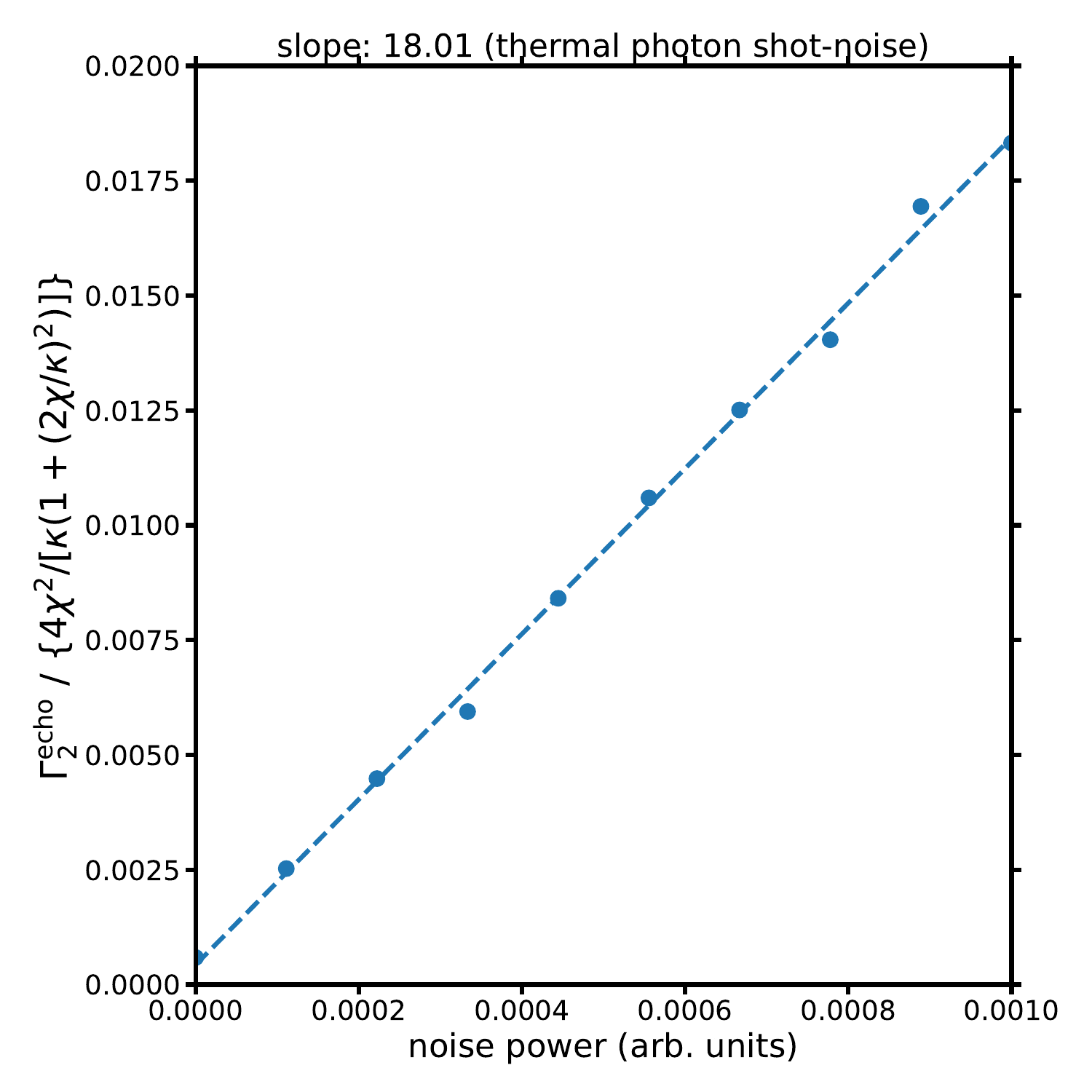}
    \includegraphics[width=0.85\columnwidth,trim=0cm 0cm 0cm 0cm,clip=True]{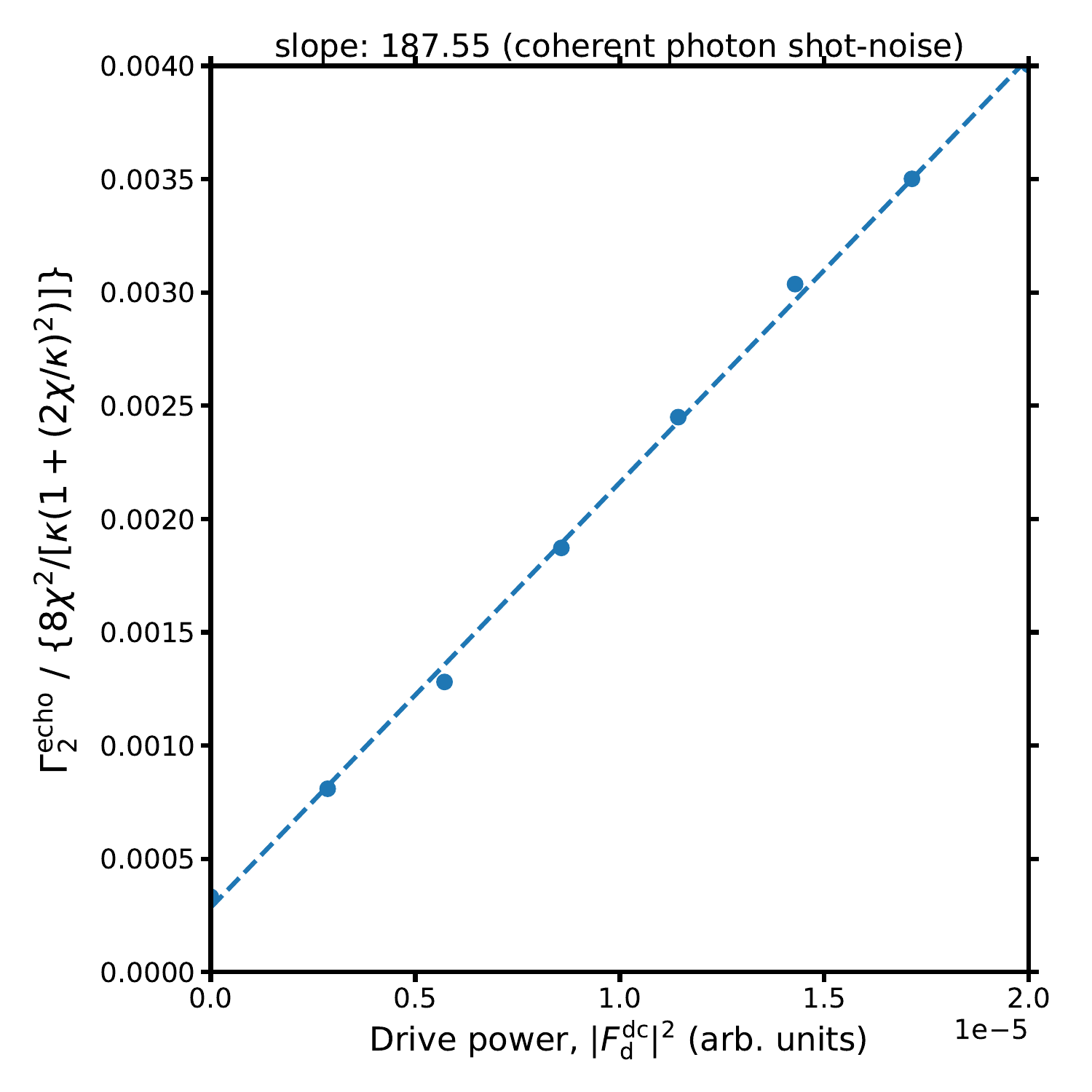}
    \caption{Spin-echo dephasing rate {\it vs.} drive power for the ``thermal'' white-noise drive (left panel) and coherent drive (right panel). These measurements are used for the  calibration of the added average photon number in the resonator.}
    \label{fig:echo_data_cal}
\end{figure*}

The low-frequency dephasing rates for the thermal and coherent cases are [see  Eqs.~(\thermalmainresultOne) and Eq.~(\cohmainresultOne) of the main text] %
\begin{align}
\label{SI:Gamma_echo_thermal_coherent}
    \Gamma_{\varphi}^{\rm th}(\Delta t\to\infty) =&\, \frac{4\chi^2\nth}{\kappa [1+(2\chi/\kappa)^2]},\nonumber \\
    \Gamma_{\varphi}^{\rm coh}(\Delta t\to \infty) =&\, \frac{8\chi^2\ncoh}{\kappa [1+(2\chi/\kappa)^2]}.
\end{align}
We used these formulas for calibration of $\nth$ and $\ncoh$. 

Figure~\ref{fig:echo_data_cal} shows the measured low-frequency dephasing rate for various drive powers $\mathcal{P}_{\rm drive}$ (in arbitrary units) for the thermal (left panel) and resonant coherent (right panel) cases. The vertical axes are scaled by the factors $4\chi/\kappa[1+(2\chi/\kappa)^2]$ and $8\chi/\kappa[1+(2\chi/\kappa)^2]$, as shown in Fig.~\ref{fig:echo_data_cal}. In the absence of additional decoherence from other noise sources not related to photon shot noise in the resonator, the vertical axes correspond to $\nth$ in the left panel and $\ncoh$ in the right panel. The sought proportionality coefficients are obtained from linear fits (depicted by the dashed lines). We find $\nth/\mathcal{P}_{\rm drive} \approx 18$ and $\ncoh/\mathcal{P}_{\rm drive} \approx 188$.  These proportionality factors (which depend on our experimental setup and arbitrary units) are then used to produce desired photon-number values $\nth$ and $\ncoh$ in the resonator.

\section{Numerical results for $\Gamma_\varphi$ with $\pi$-pulses of finite duration}
In this section we briefly discuss numerical simulations for the dephasing rate $\Gamma_\varphi$ that take into account the finite duration and shape of the experimental $\pi$-pulses. We numerically solved the following Lindblad equation for the qubit-resonator density matrix $\rho$: 
\begin{align}
\label{SI:rho_dot_numerics}
    \dot{\rho} =&\; -i[H(t) , \rho] + \mathcal{L}_{\rm res}[\rho], \nonumber\\
    H(t) =&\; -(\chi \sigma_z + \delta\omega_{\rm d})\hat{n} + \sqrt{\kappa}[F_{\rm d}^{\rm dc}a^\dagger - (F^{\rm dc}_{\rm d})^*a]i \nonumber \\
    &\; + \sum_{t_{\rm p}} g(t-t_{\rm p})\sigma_x,
\end{align}
where $t_{\rm p}$ indicates the center instant of one of the $\pi$-pulses in the CPMG sequence and $g(t) = \pi[1+\cos(2\pi t/\tau)]/2\tau$ is the shape temporal function of the $\pi$-pulses [$g(t)$ is finite only in the time interval $(-\tau/2, \tau/2)$, where $\tau$ is the duration of the $\pi$-pulses]. In Eq.~\eqref{SI:rho_dot_numerics}, $\mathcal{L}_{\rm res}[\rho]$ indicates the Lindblad operator that accounts for the resonator decay; it is given by the last two lines of Eq.~\eqref{SI:rho_01_eqn} with $\rho_{01}$ replaced by $\rho$. 

In our numerical simulations we first consider a thermalization step where we evolve $\rho$ from the initial state $|0,0\rangle\langle0,0|$ (zero photons in the resonator and the qubit in the ground state) using Eq.~\eqref{SI:rho_dot_numerics} without $\pi$-pulses ($g=0$) for a time $100/\kappa$ that is much longer than the resonator decay time. Immediately after this thermalization step, we transform the system state according to an instantaneous $\pi/2$-pulse (assumed to occur at the time $t=0$ that also indicates the start of the CPMG sequence). Then, we evolve the system state according to Eq.~\eqref{SI:rho_dot_numerics} with the $\pi$-pulses centered at the times $t_{\rm p}=\Delta t/2, 3\Delta t/2,\cdots (2N-1)\Delta t/2$ where $N$ is the number of $\pi$-pulses in the sequence, and we assume $N\geq1$. The qubit coherence is given by $\mathcal{C}(t_{\rm cpmg})=2|{\rm Tr}_{\rm res}\,\langle0|\rho(t_{\rm cpmg})|1\rangle|$, where $t_{\rm cpmg}=N\Delta t$ and the trace is over the resonator Fock states (in our simulations we consider the first 5 Fock states, which is sufficient for small  average photon numbers in the resonator, $\bar n\ll1$). To compute the dephasing rate $\Gamma_\varphi$, for a given interpulse period $\Delta t$, we run  simulations for various number $N$ of $\pi$-pulses and extract $\Gamma_\varphi$ from an exponential fit of $\mathcal{C}(N\Delta t)$, as discussed in the main text.  

To benchmark our analytical formulas for the photon-induced dephasing rate that were derived assuming instantaneous $\pi$-pulses, we perform numerical simulations based on Eq.~\eqref{SI:rho_dot_numerics} with $\pi$-pulses of duration $\tau=25$~ns (so the maximum possible CPMG frequency is $1/2\tau=20$~MHz); in these calculations we consider three values of the ratio $2\chi/\kappa=0.7$, 1.4 and 2.8. 

\begin{figure*}
    \centering
    \includegraphics[width=0.95\columnwidth,trim=0cm 0cm 0cm 0cm,clip=True]{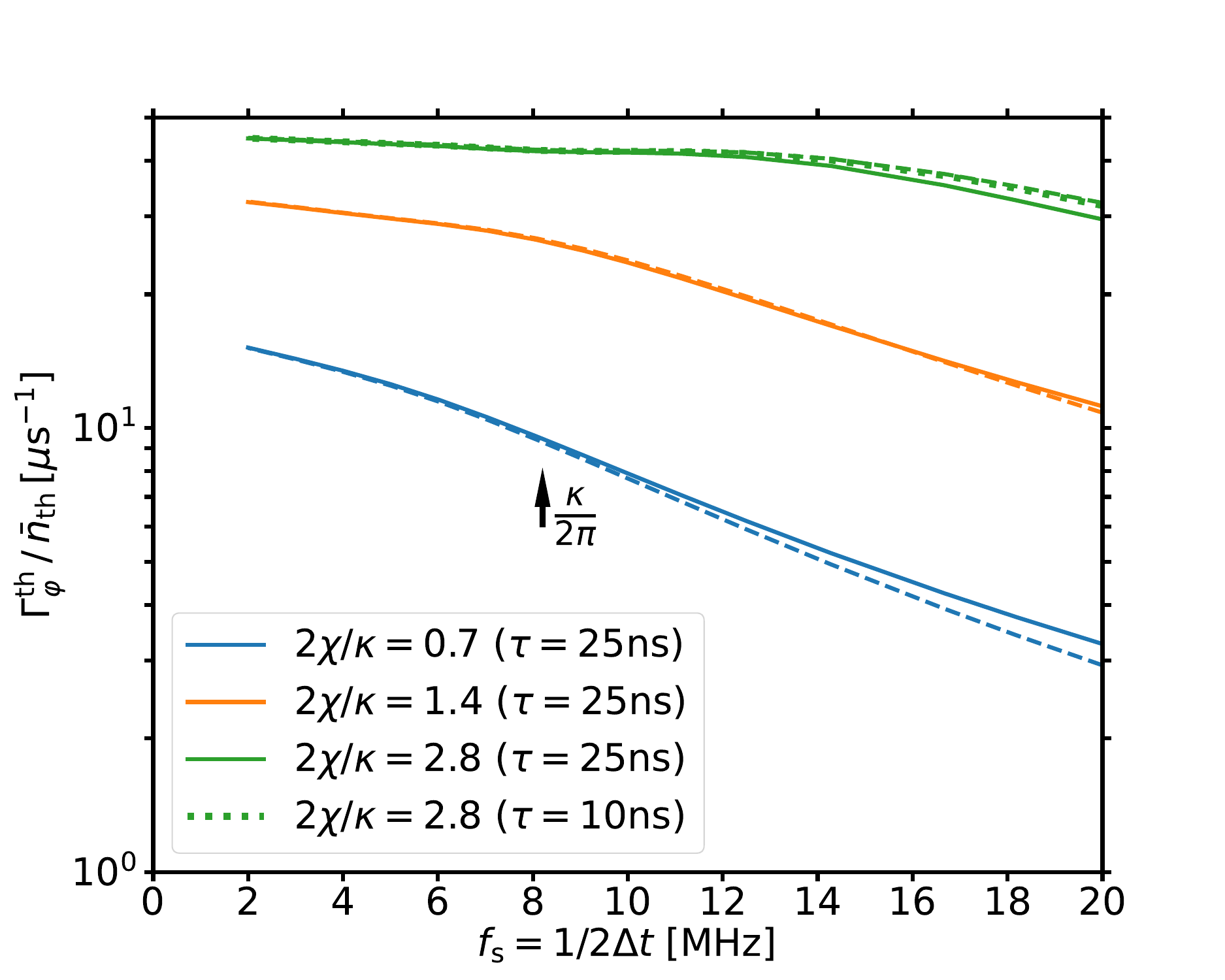}
    \includegraphics[width=0.95\columnwidth,trim=0cm 0cm 0cm 0cm,clip=True]{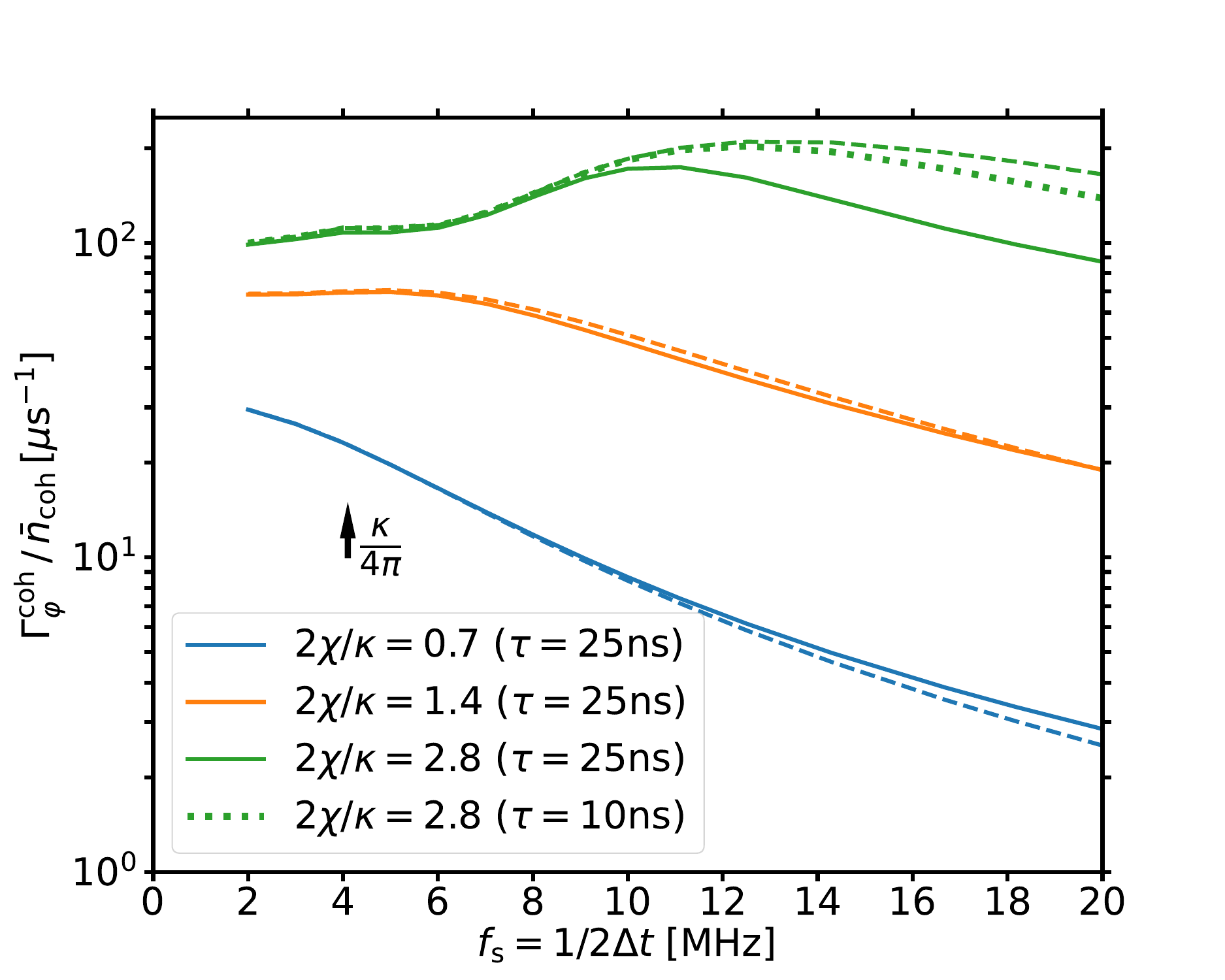}
    \caption{Dephasing rate {\it vs.} CPMG frequency for $\pi$-pulses of finite duration $\tau$. Solid lines show the simulation results that account for the shape and finite duration of the $\pi$-pulses for $\tau=25$~ns and $2\chi/\kappa=0.7$, 1.4, and 2.8. Dotted lines indicate the simulation results for $\tau=10$~ns and $2\chi/\kappa=2.8$. Dashed lines show the results from our analytical formulas (which assume $\tau=0$) given in the main text: Eqs.~(\thermalmainresultOne)--(\thermalmainresultTwo) for the case of thermal photons (left panel) and Eqs.~(\cohmainresultOne)--(\cohmainresultTwo) for the case of coherent photons in the resonator (right panel). We use $\kappa=1/(19.4\,{\rm ns})$ and $\nth=\ncoh=5\times 10^{-3}$.}
    \label{fig:finite_pi_pulses}
\end{figure*}

The left (right) panel of Fig.~\ref{fig:finite_pi_pulses} shows the photon-induced dephasing rate for the case of thermal (coherent) photons in the resonator with an average photon number of $\nth=5\times10^{-3}$ ($\ncoh=5\times10^{-3}$ and with vanishing detuning, $\delta\omega_{\rm d}=0$). In both panels, solid lines indicate the simulation results [based on Eq.~\eqref{SI:rho_dot_numerics}] that assume $\pi$-pulses of duration $\tau=25$~ns and the dashed lines indicate the results obtained from the analytical formulas for the photon-induced rate given in the main text (analytics  assume $\tau=0$). By comparing the solid and dashed lines, we find that, due to the finite duration of the $\pi$-pulses,  the discrepancy between the simulations and the analytical formulas increases as we increase the parameter $2\chi/\kappa$ and at large CPMG frequencies (in particular, the discrepancy becomes non negligible when the interpulse period is shorter than roughly twice the $\pi$-pulse duration; for $\tau=25$~ns, this corresponds to a CPMG frequency of 10~MHz in the plots of Fig.~\ref{fig:finite_pi_pulses}). We point out that, for a given $2\chi/\kappa$, the simulation results approach the results of the analytical formulas if we consider faster $\pi$-pulses (i.e., smaller $\tau$); for instance, the green dotted lines in Fig.~\ref{fig:finite_pi_pulses} (corresponding to $\tau=10$~ns and $2\chi/\kappa=2.8$) are much closer than the green solid lines that assume $\tau=25$~ns (and also $2\chi/\kappa=2.8$).



\section{Filter-function approach for CPMG}

Here we discuss the filter-function approach to find qubit dephasing in CPMG sequence and, in particular, discuss derivation of Eq.~(\Gammaphifilterfunc) of the main text. The filter-function approach is well developed (see, e.g., Refs.~\cite{S-Cappellaro2017, S-DasSarma2008, Bylander-2011, Krantz-19, Biercuk-2012}); however, there are different notations in different papers, so for completeness we review here the filter-function approach in the simple and not necessarily rigorous way. 

As in the main text, let us consider the standard single-qubit CPMG sequence containing $N$ $\pi$-pulses with period $\Delta t$ for  $\pi$-pulses, so that the total duration of the sequence is $t_{\rm cpmg}=N\Delta t$. In more detail, the sequence starts at $t=0$ with $\pi/2$-pulse over $Y$ axis, followed by $N$ $\pi$-pulses over $X$-axis at time moments  $(1/2+k)\Delta t$ with $k=0,\,1,\,...N-1$, then followed by $\pi/2$-pulse over $Y$ axis at time $N\Delta t$, which ends the sequence -- see Fig.~1 of the main text. ($N=1$ corresponds to a variant of the echo sequence.) All pulses are assumed to be instantaneous and perfect. Initial qubit state is $|0\rangle$, so after the first $\pi/2$-pulse it becomes $(|0\rangle+|1\rangle)/\sqrt{2}$, then in the ideal case $\pi$-pulses do not change this state, and after the second $\pi/2$-pulse the qubit state becomes $|1\rangle$. Note that even if the $\pi$-pulses are not exactly over $X$ axis, but still on resonance with the qubit, the state remains on the equator of the Bloch sphere; therefore with scanned phase of the second $\pi/2$-pulse, the visibility remains ideally 100\%. Thus, the visibility  shows effect of dephasing even with imperfect phases of $\pi$-pulses.

We assume no energy relaxation, $T_1=\infty$, and  discuss pure dephasing time $T_\varphi$. In reality, experimentally measured quantity is dephasing time $T_2$, from which the pure dephasing can be extracted as $1/T_\varphi =1/T_2 -1/2T_1$.

If the qubit frequency $\Omega_q (t)$ fluctuates as 
    \be
    \Omega_q(t) =\Omega_{q, 0} +\delta \Omega_q (t), 
    \ee
then the extra phase accumulated by the qubit during  total duration $N\Delta t$ is
    \be
    \varphi = \int\nolimits_0^{N\Delta t} g(t) \, \delta \Omega_q(t) \, dt , \,\,\, g(t)=\pm 1,
    \label{varphi-accum-phase}\ee
where $g(t)=1$ for $0<t<\Delta t /2$ and then flips the sign after each $\pi$-pulse.

It is easy to show that if $\delta \Omega_q (t)$ 
oscillates with a frequency $\omega$ and a complex amplitude $A$, 
$\delta \Omega_q(t) = {\rm Re} (A e^{i\omega t})$, then
$\varphi = {\rm Re} \big[ A (i/\omega) (1-e^{i\omega \Delta t/2})^2 [1- (-1)^N e^{i\omega N\Delta t}]/(1+e^{i\omega\Delta t})\big]$. Therefore, averaging over a random phase of the complex amplitude $A$, we get $\langle \varphi \rangle=0$ and $\langle \varphi^2 \rangle = (|A|^2/2\omega^2) |1-e^{i\omega \Delta t/2}|^4 |1-(-1)^N e^{i\omega N\Delta t}|^2 / |1+e^{i\omega \Delta t}|^2$. This expression can be simplified using relations $|1-e^{i\alpha} |^2= 4\sin^2 (\alpha/2)$ and $|1+e^{i\alpha} |^2= 4\cos^2 (\alpha/2)$; therefore, $|1-(-1)^N e^{i\omega N\Delta t}|^2$ becomes $\sin^2(\omega N\Delta t/2)$ for even $N$ and $\cos^2(\omega N\Delta t/2)$ for odd $N$. Also, instead of the amplitude $|A|$, let us use variance of the qubit frequency fluctuation, $\langle (\delta \Omega_q)^2\rangle=|A|^2/2$.

Next, we assume that the phase $\varphi$ is Gaussian-distributed (Gaussian noise), so that 
    \be
    \langle \cos \varphi \rangle = e^{-\langle \varphi^2\rangle /2}, 
    \ee
then the pure dephasing time $T_\varphi$ can be introduced as 
    \be
    \frac{T}{T_\varphi} = \frac{\langle \varphi^2\rangle}{2}, \,\,\,\,\, T=N\Delta t,
    \ee
so that the CPMG-sequence visibility due to pure dephasing is $e^{-T/T_\varphi}$. Thus, using the formula from the previous paragraph for $\langle \varphi^2\rangle$, we obtain for even $N$
    \be
\frac{N\Delta t}{T_\varphi} =   \langle (\delta \Omega_q)^2\rangle \, \frac{8}{\omega^2} \, \frac{\sin^4 (\omega \Delta t/4)}{\cos^2(\omega\Delta t/2)} \sin^2 (\omega N\Delta t/2),
    \label{Ndt/Tphi-1}\ee
while for odd $N$ the last term $\sin^2 (\omega N\Delta t/2)$ should be replaced with $\cos^2 (\omega N\Delta t/2)$.

Now let us integrate over fluctuations at all frequencies $\omega$, introducing the spectral density  $S_{\Omega_q} (\omega)$ of the qubit frequency fluctuations. We will use the {\it engineering} normalization of the {\it single-sided} spectral density, so that
    \be
    \langle (\delta \Omega_q)^2\rangle = \int\nolimits_0^\infty S_{\Omega_q}(\omega) \, \frac{d\omega}{2\pi} \, . 
    \label{dOmega-S}\ee
Using Eq.\ (\ref{Ndt/Tphi-1}), we obtain
\be
 \frac{N\Delta t}{T_\varphi} = \frac{\langle \varphi^2\rangle}{2} = \int\nolimits_0^\infty F(\omega) \, S_{\Omega_q}(\omega) \, \frac{d\omega}{2\pi} \, ,
\label{NDt/Tphi-2}\ee
where the filter function $F(\omega)$ for even $N$ is
    \begin{eqnarray}
&&    F(\omega) = \frac{8}{\omega^2} \, \frac{\sin^4 (\omega \Delta t/4)}{\cos^2(\omega\Delta t/2)} \sin^2 (\omega N\Delta t/2)
    \label{F-1}\\
&& \hspace{0.9cm} =  \frac{2}{\omega^2} \left[ 1-\frac{1}{\cos (\omega \Delta t/2)} \right]^2 \sin^2 (\omega N\Delta t/2), \qquad
    \label{F-2}\end{eqnarray}
while for odd $N$ the last term $\sin^2 (\omega N\Delta t/2)$ should be replaced with $\cos^2 (\omega N\Delta t/2)$ \cite{S-DasSarma2008}. Note that if the double-sided spectral density is used (which is twice smaller and symmetric for a classical noise), then the integration in Eqs.\ (\ref{dOmega-S}) and (\ref{NDt/Tphi-2}) should be extended from $-\infty$ to $\infty$, while $F(\omega)$ remains the same.

The filter functions defined in several other papers are related to our $F(\omega)$ via various coefficients. The filter function defined in Ref.\  \cite{S-Cappellaro2017} [called there $|Y(\omega)|^2$] is $F/2$ -- see, e.g., Eqs.\ (88) and (91) in  \cite{S-Cappellaro2017}. The filter function defined in Ref.\ \cite{S-DasSarma2008} (called there $F(z)$ with $z=\omega N\Delta t$) is $\omega^2 F$ -- see Table I in  \cite{S-DasSarma2008} and also Eqs.\ (17) and (18), in which $S(\omega)$ is the double-sided spectral density. The filter function defined in Refs.\  \cite{Bylander-2011, Krantz-19} (called there $g_N(\omega,\tau)$ with $\tau=T=N\Delta t$) is $(2/T^2)F$ -- see Eqs.\ (57)--(61) in \cite{Krantz-19}.

The filter function (\ref{NDt/Tphi-2})--(\ref{F-2}) satisfies the relation
    \be
 \int\nolimits_0^\infty F (\omega) \, \frac{d\omega}{2\pi} = \frac{T}{4}\, , 
    \label{F-norm}\ee
so that for a white noise, $S_{\Omega_q}(\omega)={\rm const}$, we have
    \be
\frac{1}{T_\varphi} =\frac{S_{\Omega_q}}{4},  
    \ee
independent of $N$ (CPMG does not help against white noise).

For completeness let us also mention the filter function for the Ramsey sequence: $F(\omega)=2\sin^2(\omega T/2)/\omega^2$. It also satisfies Eq.\ (\ref{F-norm}). For $N=1$ (the echo sequence) the filter function is $F(\omega)= 8 \sin^4 (\omega\Delta t/4)/\omega^2$.

For $N\gg 1$, the CPMG filter function $F(\omega)$ [Eqs.\ (\ref{F-1}) and (\ref{F-2})] has a narrow main peak centered at $\omega \Delta t/2\pi =1/2$ and bounded by zeros at $\omega \Delta t/2\pi =1/2 \pm 1/N$. The second significant peak (9 times smaller that the main peak) is in between $\omega \Delta t/2\pi =3/2 \pm 1/N$, the third peak (25 times smaller) is in between  $5/2 \pm 1/N$ and so on. There are also small additional peaks, mainly visible between the main peak and the second peak. In the vicinity of the main peak, the filter function can be approximated as
    \be
    F(\omega) \approx \frac{2N^2}{(\pi/\Delta t)^2} \, \frac{\sin^2(\delta\omega N\Delta t/2)}{(\delta\omega N\Delta t/2)^2}, \,\,\, \delta \omega =\omega- \frac{\pi}{\Delta t},
    \label{main-peak-approx}\ee
which shows small additional peaks near the main peak. From this formula, the integral of the main peak (within $1/2\pm 1/N$) is approximately 73\% of the total value (\ref{F-norm}). Including additional peaks near the main peak, this fraction is $8/\pi^2\approx 81\%$. The filter function near the second peak is given by Eq.\ (\ref{main-peak-approx}) multiplied by $1/9$ and with $\delta \omega =\omega-3\pi/\Delta t$. For the third peak the factor is $1/25$ and $\delta \omega =\omega-5\pi/\Delta t$, and so on.

Replacing Eq.\ (\ref{main-peak-approx}) with the $\delta$-function (with the same full integral) and doing the same for other peaks, we obtain a simple formula for $N\gg 1$,
    \be
    F(\omega) \approx N\Delta t \, \frac{2}{\pi^2} \sum_{m=0}^\infty \frac{1}{(2m+1)^2} \, \delta \left( \frac{\omega}{2\pi} - \frac{2m+1}{2\Delta t}\right) .
    \label{F-sum}\ee

The approximation $N\gg 1$ can be applied to the filter function $F(\omega)$ crudely starting with $N\geq 3$, while the spin-echo case $N=1$ is very different (in particular the maximum is achieved at $\omega \Delta t/2\pi \approx 0.74$ instead of $1/2$) and the case $N=2$ is still significantly different.

\vspace{0.3cm}

Combining Eq.\ (\ref{F-sum}) with Eq.\ (\ref{NDt/Tphi-2}), we find for $N\gg 1$ (see Eq.\ (13) in Ref.~\cite{S-Hirayama2011} and Eq.\ (2) in Ref.~\cite{S-Suter2011})
    \be
    \frac{1}{T_\varphi} = \frac{2}{\pi^2} \sum_{m=0}^\infty \frac{ S_{\Omega_q}(2\pi (2m+1)f_s)}{(2m+1)^2} \, , \,\,\, f_s \equiv \frac{1}{2\Delta t}\, ,
    \label{1/Tphi}\ee
where $f_s$ is the frequency of the main peak, typically associated with CPMG sequence. If the double-sided spectral density is used, then the summation over $m$ should be from $-\infty$ to $\infty$. We emphasize that Eq.\ (\ref{1/Tphi}) assumes a Gaussian noise of the qubit frequency.

\subsection*{Dephasing by photons in a resonator}

For fluctuations of {\it thermal} photons in a resonator with a small average photon number,  $\bar{n}_{\rm th}\ll 1$, the {\it single sided} spectral density of the photon number fluctuation is (e.g.,\ \cite{S-Oliver2018}) 
    \be
    S_{n}(\omega) = \frac{4\kappa \bar{n}_{\rm th}}{\kappa^2+\omega^2},
    \ee 
where $\kappa$ is the resonator energy decay rate. Note that $\int_0^\infty S_n(\omega)  d\omega/2\pi=\bar{n}_{\rm th}$ (the double-sided spectral density is twice smaller). Now assume that the resonator is coupled to a qubit, resulting in a small dispersive shift $2\chi$ (ac Stark shift per photon). Then the qubit frequency fluctuation has the single-sided spectral density 
\be
    S_{\Omega_q} (\omega) = \frac{16\chi^2 \kappa \bar{n}_{\rm th}}{\kappa^2+\omega^2}.
    \label{S-thermal}
\ee
Substituting this formula into Eq.\ (\ref{1/Tphi}), we obtain the CPMG dephasing rate 
\be
     \frac{1}{T_\varphi} = \frac{4\chi^2 \bar{n}_{\rm th}}{\kappa} \left[ 1-\frac{\tanh (\kappa \Delta t/2)}{\kappa\Delta t/2} \right] ,
    \label{Tphi-thermal} 
\ee 
which is Eq.~(\Gammaphifilterfunc) of the main text. We emphasize that this result assumes $|2 \chi| \ll \kappa$ and $\bar{n}_{\rm th}\ll 1$. If $|2\chi| \agt \kappa$, then the noise is significantly non-Gaussian, and we cannot use Eq.\ (\ref{1/Tphi}).  

In the case of an arbitrary $|2\chi|/\kappa$ (still assuming $\bar{n}_{\rm th}\ll 1$), the low-frequency dephasing rate (corresponding to $\Delta t \to\infty$ is \cite{S-Dykman1987,S-Clerk2007} $T_\varphi^{-1} = 4\chi^2 \bar{n}_{\rm th}/\{\kappa [1+(2\chi/\kappa)^2] \}$, so it is tempting to multiply the CPMG dephasing \eqref{Tphi-thermal} by the factor $[1+(2\chi/\kappa)^2]^{-1}$. However, this does not give the correct result, as can be checked by comparison with Eqs.~(\thermalmainresultOne)--(\thermalmainresultTwo) of the main text. 
In fact, while such a modification of Eq.~\eqref{Tphi-thermal} gives a better approximation for large $\Delta t$, Eq.~\eqref{Tphi-thermal} without the modification gives a better approximation for small $\Delta t$. This is why in Eq.~(\Gammaphifilterfunc) of the main text we show the result Eq.~\eqref{Tphi-thermal} without the modification.

For the noise due to fluctuation of photon number in a resonator {\it driven on resonance}, we have (see Eqs.\ (E10) and (E11) in \cite{Clerk-2010})
    \be
    S_{\Omega_q} (\omega) = \frac{8\chi^2 \kappa \bar{n}_{\rm coh}}{(\kappa/2)^2+\omega^2},
    \label{S-Omega-drive}\ee
where $\bar{n}_{\rm coh}$ is the average number of photons (this equation assumes zero temperature). For this noise, Eq.\ (\ref{1/Tphi}) gives
    \be
    \frac{1}{T_\varphi} = \frac{8\chi^2 \bar{n}_{\rm coh}}{\kappa} \left[ 1-\frac{\tanh (\kappa\Delta t/4)}{\kappa \Delta t/4} \right] .
    \label{Tphi-driven}\ee
Note that besides assumptions of resonant drive and $|2\chi| \ll \kappa$, this formula needs sufficiently small $\bar{n}_{\rm coh}$, so that $T_{\varphi} \gg \max (\Delta t, 1/\kappa )$. Compared with Eq.\ (\ref{Tphi-thermal}), for the same intracavity photon number, dephasing rate $T_\varphi^{-1}$ is twice bigger at small frequency  $f_s= 1/2\Delta t$, but the  frequency range is twice smaller. 

\subsection*{Gaussian vs. non-Gaussian noise}

Let us discuss why in the case $|2\chi |/\kappa \ll 1$ the qubit frequency noise is Gaussian (and therefore the filter-function approach can be used), while when $|2\chi |/\kappa \agt 1$, the noise is significantly non-Gaussian (and the filter-function approach is not applicable). 

Let us start with discussing the thermal noise, assuming $\bar{n}_{\rm th}\ll 1$.  The Lindblad equation implies that a photon is created in the  resonator with the rate $\bar{n}_{\rm th} \kappa$, and then it remains for time $t_{\rm ph}$, which is exponentially distributed with average $1/\kappa$ (we assume $\bar{n}_{\rm th}\ll 1$ and neglect possibility of two and more photons). The photon creation occurs rarely, but when it occurs, the qubit phase is changed by $2\chi t_{\rm ph} \sim 2\chi /\kappa$. If  $|2\chi |/\kappa \ll 1$, then the phase change from one photon is small, and therefore many photon-creation events and decays should take place before a noticeable qubit dephasing. In this case the Central Limiting Theorem ensures that the distribution of the accumulated phase $\varphi$ in Eq.\ (\ref{varphi-accum-phase}) is Gaussian, which is needed for the filter-function approach to be applicable. However, if $|2\chi |/\kappa \agt 1$, then a single photon-creation event is sufficient to produce $|\varphi | \agt 1$ (at least if $\Delta t \agt 1/\kappa$), so the distribution of $\varphi$ is obviously non-Gaussian, and therefore the filter-function approach is not applicable.

Now let us consider the case of a resonant coherent drive (at zero temperature) with $\bar{n}_{\rm coh} \ll 1$ (this condition is necessary because otherwise for a moderate value of $|2\chi| /\kappa$  the qubit  decoheres within $1/\kappa$). The noise is now due to vacuum fluctuations of the field, so it cannot be visualized classically. Nevertheless, we can still qualitatively think in terms of discrete photon numbers or in terms of rare field kicks with $|\Delta \alpha| \sim 1$, which is a typical amplitude of quantum fluctuations. In both cases the qubit phase change due to ``elementary'' fluctuation is on the order of $2\chi/\kappa$. Then Gaussianity of the noise requires $|2\chi/\kappa| \ll 1$, similar to the reasoning in the previous paragraph.


\end{document}